\documentclass[fleqn,a4paper,printer]{aa}

\usepackage{amsmath}
\usepackage{verbatim}
\usepackage{graphicx}
\usepackage{txfonts}
\usepackage{subfigure}
\usepackage{float}
\usepackage{natbib}
\usepackage{twoopt}
\usepackage{xspace}
\usepackage{balance}
\usepackage{upgreek}
\usepackage{multirow}

\def\apjl{Astrophys. J. Lett.}
\def\apjs{Astrophys. J. Suppl.}

\def\aap{Astron. Astrophys.}

\newcommand{\Mpch}{$h^{-1}\,\mbox{Mpc}$\,}

\newcommand{\xiiz}{$\xi(s_\perp,s_\parallel)$\,}
\newcommand{\Om}{$\Omega_{\rm M}$\,}

\bibpunct{(}{)}{;}{a}{}{,} 
\DeclareGraphicsExtensions{.eps,.ps,.pdf}

\begin{document}

\title {Redshift-space distortions of galaxies, clusters and AGN}

\subtitle{Testing how the accuracy of growth rate measurements depends
  on scales and sample selections}

\author{
  Federico Marulli\inst{\ref{1},\ref{2},\ref{3}}
  \and Alfonso Veropalumbo\inst{\ref{1}}
  \and Lauro Moscardini\inst{\ref{1},\ref{2},\ref{3}}
  \and Andrea Cimatti\inst{\ref{1}}
  \and Klaus Dolag\inst{\ref{4},\ref{5}}
}

\offprints{F. Marulli \\ \email{federico.marulli3@unibo.it}}

\institute{
  Dipartimento di Fisica e Astronomia - Universit\`{a} di Bologna, viale Berti Pichat 6/2, I-40127 Bologna, Italy \label{1}
  \and INAF - Osservatorio Astronomico di Bologna, via Ranzani 1, I-40127 Bologna, Italy \label{2}
  \and INFN - Sezione di Bologna, viale Berti Pichat 6/2, I-40127 Bologna, Italy \label{3}
  \and Department of Physics, Ludwig-Maximilians-Universit\"at, Scheinerstr. 1, D-81679 M\"unchen, Germany \label{4}
  \and Max-Planck-Institut f\"ur Astrophysik, Karl-Schwarzschild Strasse 1, D-85748 Garching bei M\"unchen, Germany \label{5}
}

\abstract
% context heading 
{}
% aims heading
{Redshift-space clustering anisotropies caused by cosmic peculiar
  velocities provide a powerful probe to test the gravity theory on
  large scales. However, to extract unbiased physical constraints, the
  clustering pattern has to be modelled accurately, taking into
  account the effects of non-linear dynamics at small scales, and
  properly describing the link between the selected cosmic tracers and
  the underlying dark matter field.}
% methods heading 
{We use a large hydrodynamic simulation to investigate how the
  systematic error on the linear growth rate, $f$, caused by model
  uncertainties, depends on sample selections and comoving
  scales. Specifically, we measure the redshift-space two-point
  correlation function of mock samples of galaxies, galaxy clusters
  and Active Galactic Nuclei, extracted from the {{\em Magneticum}}
  simulation, in the redshift range $0.2\leq z\leq2$, and adopting
  different sample selections. We estimate $f\sigma_8$ by modelling
  both the monopole and the full two-dimensional anisotropic
  clustering, using the {\em dispersion model}.}
% results heading 
{We find that the systematic error on $f\sigma_8$ depends
  significantly on the range of scales considered for the fit. If the
  latter is kept fixed, the error depends on both redshift and sample
  selection, due to the scale-dependent impact of non-linearities, if
  not properly modelled. On the other hand, we show that it is
  possible to get almost unbiased constraints on $f\sigma_8$ provided
  that the analysis is restricted to a proper range of scales, that
  depends non trivially on the properties of the sample. This can have
  a strong impact on multiple tracers analyses, and when combining
  catalogues selected at different redshifts.}
% conclusions heading 
{}


\authorrunning{F. Marulli et al.}
\titlerunning{Redshift-space distortions with different tracers}

\maketitle

%%%%%%%%%%%%%%%%%%%%%%%%%%%%%%%%%%%%%%%%%%%%%%%%%%%%%%%%%%%%%%%%%%%%%%%%%%%%%%%
%%%%%%%%%%%%%%%%%%%%%%%%%%%%%%%%%%%%%%%%%%%%%%%%%%%%%%%%%%%%%%%%%%%%%%%%%%%%%%%

\section{Introduction}

The dynamics of the Universe is shaped by the gravity force. The
understanding of gravity is thus the key to unveil the nature of the
Universe and its cosmological evolution. After almost one century from
its original formulation, Einstein's General Relativity (GR) is still
the dominant theory to describe gravity. However, despite its lasting
experimental successes, some fundamental open issues are motivating
challenging efforts aimed at searching for possible expansions or
radically alternative models \citep{amendola2013}. First of all, GR is
not a quantum field theory and cannot be reconciled with the
principles of quantum mechanics. Secondly, an increasingly large
amount of astronomical observations cannot be explained by GR and
baryonic matter alone, requiring the introduction of a never observed
form of dark matter \citep[DM,][]{zwicky1937}. Finally, independent
cosmological observations give indisputable evidences for an
accelerated expansion of the Universe, requiring the introduction of
another dark component, generally dubbed dark energy (DE), when
interpreted in the GR framework \citep{riess1998,
  perlmutter1999}. Whether GR provides a reliable description of the
large scale structure of the Universe, and thus the latter is
dominated by unknown dark components, or, on the contrary, such a
gravity theory has to be corrected, or even abandoned, is one of the
key questions of modern physics and cosmology.

One of the most effective ways to test the gravity theory on large
scales, that is where DM and DE arise, is to exploit the apparent
anisotropies observed in galaxy maps, the so-called redshift-space
distortions \citep[RSD,][]{kaiser1987, hamilton1998}.  Indeed, by
modelling redshift-space clustering anisotropies it is possible to
constrain the linear growth rate of cosmic structures, $f\equiv
d\log\delta/d\log a$, where $a$ is the dimensionless scale factor and
$\delta$ the linear fractional density perturbation, provided that the
galaxy bias is known.  The first RSD measurements were exploited
primarily to estimate the mean matter density, \Om, that can be
derived directly from $f(z)$ when a gravity model is assumed
\citep[][]{peacock2001, hawkins2003}. Later on, \citet{guzzo2008} and
\citet{zhang2008} showed that RSD can be effectively used to
discriminate between DE and modified gravity scenarios. RSD thus
started to be considered as a powerful probe of gravity, and several
applications to galaxy redshift surveys followed rapidly, both in the
local Universe and at larger redshifts, up to $z\sim1$, such as 6dFGS
\citep{beutler2012}, SDSS \citep{samushia2012, chuang2013,
  chuang2013b}, WiggleZ \citep{blake2012, contreras2013}, VIPERS
\citep{delatorre2013b}, and BOSS \citep{tojeiro2012, reid2012,
  reid2014}.

All the above measurements have been obtained from large galaxy
redshift surveys. First tentative studies are starting to consider
also different tracers, but results are still strongly affected by
statistical uncertainties, due to the paucity of the catalogues
used. Thanks to the wealth of observational data now available, or
expected in the next future, for instance from the ESA Euclid
mission\footnote{http://www.euclid-ec.org} \citep{laureijs2011}, the
NASA Wide Field Infrared Space Telescope (WFIRST)
mission\footnote{http://wfirst.gsfc.nasa.gov} \citep{spergel2013}, and
the extended Roentgen Survey with an Imaging Telescope Array (eROSITA)
satellite mission\footnote{http://www.mpe.mpg.de/eROSITA}
\citep{merloni2012}, it will soon become possible to apply these
methodologies on both cluster and Active Galactic Nuclei (AGN)
catalogues. Galaxy clusters are the biggest collapsed structures in
the Universe. They appear more strongly biased than galaxies, that is
their clustering signal is higher. Moreover, they are relatively
unaffected by non-linear dynamics on small scales. This could help in
modelling redshift-space distortions in their clustering pattern, as
the effect of small scale incoherent motions, that generate the
so-called fingers-of-God pattern, is substantially less severe
relative to the galaxy clustering case, improving also the
cosmological constraints that can be extracted from baryon acoustic
oscillations \citep{veropalumbo2014}. The main drawback is the low
density of galaxy cluster samples and the fact that these sources can
be reliably detected only at relatively small redshifts. On the other
hand, AGN can be detected up to very large distances. Powered by
accreting supermassive black holes hosted at their centres, their
extreme luminosities make them optimal tracers to investigate the
largest scales of the Universe, and thus its cosmological evolution
\citep[see e.g.][and references therein]{marulli2008, marulli2009,
  bonoli2009}.

The aim of this paper is to investigate the {\em accuracy} of the
so-called {\em dispersion model} (that will be described in
\S\ref{sec:RSD}) as a function of comoving scales and sample
selection. To achieve this goal, we exploit realistic mock samples of
galaxies, galaxy clusters and AGN extracted from the {\em Magneticum}
hydrodynamic simulation, at six snapshots in the range $0.2\leq
z\leq2$. We adopt a similar methodology as in \citet{bianchi2012},
though substantially extending that analysis. Specifically, i) in line
with the majority of recent studies, we investigate the systematic
errors of $f\sigma_8$, instead of $\beta$, ii) we consider different
kinds of realistic mock tracers, instead of just DM haloes, iii) we
analyse the dependency of the errors as a function of redshift, and
iv) as a function of different sample selections.  In agreement with
previous investigations \citep[see e.g.][]{bianchi2012, mohammad2016},
we find that the rough modelisation of non-linear dynamics provided by
the dispersion model introduces systematic errors on $f\sigma_8$
measurements, that depend non trivially on the redshift and bias of
the tracers. On the other hand, as we will demonstrate, it is possible
to substantially reduce the systematic errors on $f\sigma_8$ provided
that the fit is restricted in a proper range of scales, that depends
on sample selection. An alternative investigation of the impact of the
galaxy sample selection function on the ability to test GR with RSD
has been carried out recently by \citet{hearin2015}, analysing the
small-scale effects of the assembly bias in velocity space.

The software that we implemented to perform all the analyses presented
in this paper can be freely downloaded at this link:
http://apps.difa.unibo.it/files/people/federico.marulli3/. It consists
of a set of C++ libraries that can be used for several astronomical
calculations, in particular to measure the two-point correlation
function and to model RSD \citep[{\small
    CosmoBolognaLib},][]{marulli2016}. We also provide the full
documentation for these libraries, and some example codes that explain
how to use them.

The paper is organised as follows. In \S\ref{sec:magneticum} we
describe the {\em Magneticum} simulation used to construct mock
samples of galaxies, clusters and AGN, whose main properties are
reported in Appendix \ref{app:samples}. In \S\ref{sec:RSD} and
\S\ref{sec:methodology} we present the formalism used to model RSD and
to measure the two-point correlation function in redshift-space mock
catalogues, respectively. The results of our analyses are presented in
\S\ref{sec:results}. In \S\ref{sec:conclusions} we summarise our
conclusions. Finally, in the appendices \ref{app:random} and
\ref{app:errors} we discuss the tests performed to investigate the
robustness of the methods applied in our analysis.

%%%%%%%%%%%%%%%%%%%%%%%%%%%%%%%%%%%%%%%%%%%%%%%%%%%%%%%%%%%%%%%%%%%%%%%%%%%%%%%

\section{The {\em Magneticum} simulation}
\label{sec:magneticum}

The most direct way to investigate systematic errors on linear growth
rate measurements from RSD is to exploit large numerical simulations,
and to apply on them the same methodologies used for real data. As our
goal is to study the effects of different sample selections, we make
use of realistic mock catalogues of different cosmic
tracers. Specifically, we analyse galaxy, cluster and AGN mock
catalogues extracted from the hydrodynamic {\em Magneticum}
simulation\footnote{www.magneticum.org} \citep{dolag2016}.

This simulation is based on the parallel cosmological TreePM-SPH code
{\small P-GADGET3} (\citealp{springel2005gadget2}). The code uses an
entropy-conserving formulation of smoothed-particle hydrodynamics
(SPH) \citep{springel2002} and follows the gas using a low-viscosity
SPH scheme to properly track turbulence \citep{dolag2005}.  It also
allows a treatment of radiative cooling, heating from a uniform
time-dependent ultraviolet background and star formation with the
associated feedback processes. The latter is based on a sub-resolution
model for the multiphase structure of the interstellar medium (ISM)
\citep{springel2003b}.

Radiative cooling rates are computed by following the same procedure
presented by \citet{wiersma2009}. We account for the presence of the
cosmic microwave background and of ultraviolet/X-ray background
radiation from quasars and galaxies, as computed by
\citet{haardt2001}. The contributions to cooling from each one of the
following 11 elements, H, He, C, N, O, Ne, Mg, Si, S, Ca, Fe, have
been pre-computed using the publicly available {\small CLOUDY}
photoionization code \citep{ferland1998} for an optically thin gas in
photoionization equilibrium.

In the multiphase model for star-formation \citep{springel2003}, the
ISM is treated as a two-phase medium where clouds of cold gas form
from cooling of hot gas and are embedded in the hot gas phase assuming
pressure equilibrium whenever gas particles are above a given
threshold density. The hot gas within the multiphase model is heated
by supernovae and can evaporate the cold clouds. A certain fraction of
massive stars (10 per cent) is assumed to explode as supernovae type
II (SNII). The released energy by SNII ($10^{51}$~erg) is modelled to
trigger galactic winds with a mass loading rate being proportional to
the star formation rate to obtain a resulting wind velocity of
$v_{\mathrm{wind}} = 350$ km/s.  Our simulation also includes a
detailed model of chemical evolution according to
\citet{tornatore2007}. Metals are produced by SNII, by supernovae type
Ia (SNIa) and by intermediate and low-mass stars in the asymptotic
giant branch (AGB). Metals and energy are released by stars of
different mass by properly accounting for mass-dependent lifetimes,
with a lifetime function according to \citet{padovani1993}, the
metallicity-dependent stellar yields by \citet{woosley1995} for SNII,
the yields by \citet{vandenhoek1997} for AGB stars and the yields by
\citet{thielemann2003} for SNIa. Stars of different mass are initially
distributed according to a Chabrier initial mass function
\citep{chabrier2003}.

Most importantly, the {\em Magneticum} simulation also includes a
prescription for black hole growth and for feedback from AGN.  As for
star formation, the accretion onto black holes and the associated
feedback adopts a sub-resolution model. Black holes are represented by
collisionless ``sink particles'' that can grow in mass by accreting
gas from their environments, or by merging with other black holes.
Our implementation is based on the model presented in
\citet{springel2005} and \citet{dimatteo2005}, including the same
modifications as in the study of \citet{fabjan2010} and some new,
minor changes, as described in \citet{hirschmann2014}.

The black hole (BH) gas accretion rate, $\dot{M}_{\rm BH}$, is
estimated by using the Bondi-Hoyle-Lyttleton approximation
\citep{hoyle1939, bondi1944, bondi1952}:
\begin{equation}
  \dot{M}_{\rm BH} = \frac{4 \pi G^2 M_{\rm BH}^2 f_{\rm boost}
    \rho}{(c_s^2 + v^2)^{3/2}},
  \label{Bondi}
\end{equation}
where $\rho$ and $c_s$ are the density and the sound speed of the
surrounding (ISM) gas, respectively, $f_{\rm boost}$ is a boost
factor for the density which is typically set to $100$ and $v$ is
the velocity of the black hole relative to the surrounding gas. The
black hole accretion is always limited to the Eddington rate. The
radiated bolometric luminosity, $L_{\rm bol}$, is related to the
black hole accretion rate by
\begin{equation}
  L_{\rm bol} = \epsilon_{\rm r} \dot{M}_{\rm BH} c^2,
\end{equation} 
where $\epsilon_{\rm r}$ is the radiative efficiency, for which we
adopt a fixed value of $0.1$, standardly assumed for a radiatively
efficient accretion disk onto a non-rapidly spinning black hole
according to \citet{shakura1973} \citep[see
  also][]{springel2005gadget2, dimatteo2005}.

The simulation covers a cosmological volume with periodic boundary
conditions initially occupied by an equal number of $1526^3$ gas and
DM particles, with relative masses that reflect the global baryon
fraction, $\Omega_{\rm b}/$\Om. The box side is $896$ \Mpch.  The
cosmological model adopted is a spatially flat $\Lambda$CDM universe
with matter density \Om$=0.272$, baryon density $\Omega_{\rm
  b}=0.0456$, power spectrum normalisation $\sigma_8=0.809$, and
Hubble constant $H_0=70.4 {\rm km\, s^{-1} Mpc^{-1}}$, chosen to match
the seven-year Wilkinson Microwave Anisotropy Probe \citep[WMPA7,
][]{komatsu2011}.  More details on the {\em Magneticum} simulation and
the derived mock catalogues can be found in \citet{hirschmann2014,
  saro2014, bocquet2015}.

Firstly, we will analyse large samples of mock galaxies, clusters and
AGN selected in stellar mass, cluster mass and BH mass,
respectively. Then, we will consider several subsamples with different
selections, as reported in Tables \ref{tab:table1}, \ref{tab:table2}
and \ref{tab:table3} (see \S\ref{sec:results}).

%%%%%%%%%%%%%%%%%%%%%%%%%%%%%%%%%%%%%%%%%%%%%%%%%%%%%%%%%%%%%%%%%%%%%%%%%%%%%%% 

\section{Redshift-space distortions}
\label{sec:RSD}

To construct redshift-space mock catalogues, we adopt the same
methodology used in \citet{bianchi2012} and \citet{marulli2011,
  marulli2012a, marulli2012b}. We provide here a brief description. We
consider a local virtual observer at $z=0$, and place the centre of
each snapshot of the {\em Magneticum} simulation at a comoving
distance, $D_{\rm c}$, corresponding to its redshift, that is:
\begin{equation}
D_{\rm c}(z)=c\int_0^z\frac{dz_c'}{H(z_c')} \; ,
\label{eq:distance}
\end{equation} 
where $c$ is the speed of light, and the Hubble expansion rate is:
\begin{multline}
H(z) = H_0\left[\Omega_{\rm M}(1+z)^3+\Omega_{\rm
    k}(1+z)^2\right. \\ \left. +\Omega_{\rm DE}
  \exp\left(3\int_0^z\frac{1+w(z)}{1+z}\right)\right]^{0.5} \; ,
\label{eq:HubbleFull}
\end{multline}
where $\Omega_{\rm k}=1-\Omega_{\rm M}-\Omega_{\rm DE}$, $w(z)$ is the
DE equation of state, and the contribution of radiation is assumed
negligible. In our case: $\Omega_{\rm k}=1$, $w(z)=1$, so $\Omega_{\rm
  DE}\equiv\Omega_{\Lambda}=1-\Omega_{\rm M}$ and the Hubble expansion
rate reduces to:
\begin{equation}
H(z) = H_0\left[\Omega_{\rm M}(1+z)^3+\Omega_{\Lambda}\right] \; .
\label{eq:Hubble}
\end{equation}

To estimate the distribution of mock sources in redshift-space, we
then transform the comoving coordinates of each object into angular
positions and observed redshifts. The latter are computed as follows:
\begin{equation}
z_{\rm obs}=z_{\rm c}+\frac{v_\parallel}{c}(1+z_{\rm c}) \; ,
\label{eq:redshift}
\end{equation}
where $z_{\rm c}$ is the {\em cosmological} redshift due to the Hubble
recession velocity at the comoving distance of the object and
$v_\parallel$ is the line-of-sight component of its centre of mass
velocity. In this analysis we do not include either errors in the
observed redshift, or geometric distortions caused by an incorrect
assumption of the background cosmology \citep[see][for more
  details]{marulli2012b}.

In the linear regime, the velocity field can be determined from the
density field, and the amplitude of RSD is proportional to the
parameter $\beta$, defined as follows:
\begin{equation}
  \beta(z) \equiv \frac{f(z)}{b(z)} \; ,
\label{eq:beta}
\end{equation}
where the linear growth rate, $f$, can be approximated in most
cosmological frameworks as:
\begin{equation}
  f(z) \simeq \Omega_{\rm M}^{\gamma}(z) \; ,
\label{eq:fs8}
\end{equation}
with $\gamma\simeq0.545$ in $\Lambda$CDM. The linear bias factor can
be estimated as:
\begin{equation}
  b(z) = \langle \sqrt{\frac{\xi(r;z)}{\xi_{\rm DM}(r;z)}} \rangle \;
  ,
\label{eq:bias}
\end{equation}
where the mean is computed at sufficiently large scales at which
non-linear effects can be neglected.

In the linear regime, the redshift-space two-point correlation
function can be written as follows:
\begin{equation} 
\xi^{\rm lin}(s,\mu) =
\xi_0(s)P_0(\mu)+\xi_2(s)P_2(\mu)+\xi_4(s)P_4(\mu) \; ,
\label{eq:ximodellin}
\end{equation}
where $\mu\equiv\cos\theta=s_\parallel/s$ is the cosine of the angle
between the separation vector and the line of sight,
$s=\sqrt{s_\perp^2+s_\parallel^2}$, and $P_l$ are the Legendre
polynomials \citep{kaiser1987, lilje1989, mcgill1990, hamilton1992,
  fisher1994}. Eq.~(\ref{eq:ximodellin}) is derived in the
distant-observer approximation, that is reasonable at the scales
considered in this analysis. The multipoles of \xiiz are:
\begin{subequations}
  \begin{align}
    \xi_0(s) & = \left(1+ \frac{2}{3}\beta + \frac{1}{5}\beta^2 \right)
    \cdot \xi(r)
    \label{eq:xi0_1} \\ 
    & =  \left[ (b\sigma_8)^2 + \frac{2}{3} f\sigma_8 \cdot b\sigma_8 +
      \frac{1}{5}(f\sigma_8)^2 \right] \cdot \frac{\xi_{\rm DM}(r)}{\sigma_8^2} \; ,
    \label{eq:xi0_2} 
  \end{align}
\end{subequations}

\begin{subequations} 
  \begin{align}
    \xi_2(s) &  = \left(\frac{4}{3}\beta +
    \frac{4}{7}\beta^2\right)\left[\xi(r)-\overline{\xi}(r)\right] 
    \label{eq:xi2_1} \\ 
    & = \left[\frac{4}{3}f\sigma_8 \cdot b\sigma_8 +
      \frac{4}{7}(f\sigma_8)^2\right]\left[\frac{\xi_{\rm
          DM}(r)}{\sigma_8^2}-\frac{\overline{\xi}_{\rm
          DM}(r)}{\sigma_8^2}\right] \; ,
    \label{eq:xi2_2} 
  \end{align}
\end{subequations}

\begin{subequations} 
  \begin{align}
    \xi_4(s) & = \frac{8}{35}\beta^2\left[\xi(r) +
      \frac{5}{2}\overline{\xi}(r)
      -\frac{7}{2}\overline{\overline{\xi}}(r)\right] 
    \label{eq:xi4_1} \\ 
    & = \frac{8}{35}(f\sigma_8)^2\left[ \frac{\xi_{\rm
          DM}(r)}{\sigma_8^2} + \frac{5}{2}\frac{\overline{\xi}_{\rm
          DM}(r)}{\sigma_8^2}
      -\frac{7}{2}\frac{\overline{\overline{\xi}}_{\rm DM}(r)}{\sigma_8^2}
      \right] \; ,
    \label{eq:xi4_2} 
  \end{align}
\end{subequations}
where $\xi(r)$ and $\xi_{\rm DM}(r)$ are the real-space {\em
  undistorted} correlation functions of tracers and DM, respectively,
whereas the {\em barred} functions are:
\begin{equation} 
\overline{\xi}_{\rm DM}(r) \equiv \frac{3}{r^3}\int^r_0dr'\xi_{\rm
  DM}(r')r'{^2} \; ,
\label{eq:xi_}
\end{equation}
\begin{equation}
\overline{\overline{\xi}}_{\rm DM}(r) \equiv \frac{5}{r^5}\int^r_0dr'\xi_{\rm DM}(r')r'{^4} \; .
\label{eq:xi__} 
\end{equation}
Eqs.~(\ref{eq:xi0_2}), (\ref{eq:xi2_2}) and (\ref{eq:xi4_2}) are
derived from Eqs.~(\ref{eq:xi0_1}), (\ref{eq:xi2_1}) and
(\ref{eq:xi4_1}) respectively, using Eq.~(\ref{eq:beta}). In this
analysis, $\xi_{\rm DM}(r)$ is estimated by Fourier transforming the
linear power spectrum computed with the software {\small CAMB}
\citep{lewis2002} for the cosmological model here considered. We could
alternatively measure the DM correlation function directly from the
snapshots of the simulation. However, by using {\small CAMB} we can
substantially reduce the computational time, we get a smooth model
with no errors due to measurements and, more importantly, we can
closely mimic the analysis we would have done with real data.

Eq.~(\ref{eq:ximodellin}) is an effective description of RSD only at
large scales, where non-linear effects are negligible \citep[but see
  e.g.][for a more accurate modelling]{scoccimarro2004, taruya2010,
  seljak2011, wang2014}.  An empirical model that can account for both
linear and non-linear dynamics is the so-called dispersion model
\citep{peacock1996, peebles1980, davis1983}, that describes the
redshift-space correlation function as a convolution of the
linearly-distorted correlation with the distribution function of
pairwise velocities, $f(v)$:
\begin{equation} 
 \xi(s_\perp, s_\parallel) = \int^{\infty}_{-\infty}dv
 f(v)\xi\left(s_\perp, s_\parallel - \frac{v(1+z)}{H(z)}\right)_{\rm
   lin} \; ,
\label{eq:ximodel}
\end{equation}
where the pairwise velocity $v$ is expressed in physical coordinates.

Since we are not considering redshift errors, we use the exponential
form for $f(v)$ \citep{marulli2012b}, namely:
\begin{equation}
f(v)=\frac{1}{\sigma_{12}\sqrt{2}}
\exp\left(-\frac{\sqrt{2}|v|}{\sigma_{12}}\right)
\label{eq:fvexp} 
\end{equation}
\citep{davis1983, fisher1994, zurek1994}.  The quantity $\sigma_{12}$
can be interpreted as the dispersion in the pairwise random peculiar
velocities, and is assumed to be independent of pair separations. 

The dispersion model given by
Eqs.~(\ref{eq:ximodellin})-(\ref{eq:fvexp}) depends on three free
quantities, $f\sigma_8$, $b\sigma_8$ and $\sigma_{12}$ (since
$\xi_{\rm DM} \propto \sigma_8^2$), and on the reference background
cosmology used both to convert angles and redshifts into distances and
to estimate the real-space DM two-point correlation function.  This
model has been widely used in the past years to estimate the linear
growth rate both in configuration space, that is by modelling \xiiz or
its multipoles \citep[e.g.][]{peacock2001, hawkins2003, guzzo2008,
  ross2007, cabre2009a, cabre2009b, contreras2013} and in Fourier
space, by modelling the power spectrum \citep[e.g.][]{percival2004,
  tegmark2004, blake2011c}. In this analysis, we investigate the
accuracy of the dispersion model in configuration space, while
alternative RSD models will be analysed in an upcoming paper.

%%%%%%%%%%%%%%%%%%%%%%%%%%%%%%%%%%%%%%%%%%%%%%%%%%%%%%%%%%%%%%%%%%%%%%%%%%%%%%%

\section{Methodology}
\label{sec:methodology}

The aim of this work is to test widely-used statistical methodos to
model RSD and to extract constraints on the linear growth rate. Thus,
we will use methodologies that can be applied directly to real
data. On the other hand, we will not consider any specific mock
catalogue, in order to keep our analysis general, that is not
restricted to any specific real survey.  To measure the two-point
correlation functions of our mock samples we make use of the
\citet{landy1993} estimator:
\begin{equation}
\xi(r)=\frac{OO(r)-2OR(r)+RR(r)}{RR(r)} \; ,
\label{eqn:landy}
\end{equation}
where $OO(r)$, $OR(r)$ and $RR(r)$ are the fractions of
object--object, object--random and random--random pairs, with spatial
separation $r$, in the range $[r-\delta r/2, r+\delta r/2]$, where
$\delta r$ is the bin size.  The random catalogues are constructed to
be three times larger than the associated mock samples. As we are
considering mock catalogues in cubic boxes, with no geometrical
selection effects and periodic conditions, we could easily estimate
the two-point correlation function directly from the density field, or
computing the random counts analytically. Nevertheless, we prefer to
use the \citet{landy1993} estimator to mimic more closely the analysis
on real data, as stressed before. In any case, this choice does not
impact the results of our analysis. Moreover, as we verified, the
number of random objects used in this work is large enough to have no
significant effects on our measurements (see Appendix
\ref{app:random}).  We compute the correlation functions up to a
maximum scale of $r=50$ \Mpch, both in the parallel and perpendicular
directions. As we have verified with a sub-set of mocks, going to
higher scales does not change our results, as the constraints on
$f\sigma_8$ are mainly determined by the RSD signal at smaller scales.

To model RSD and extract constraints on $f\sigma_8$, we exploit the
model given by Eqs.~(\ref{eq:ximodellin})-(\ref{eq:fvexp}). We
consider two case studies. The first one consists in modelling only
the monopole of the redshift-space two-point correlation function at
large scales, via Eq.~(\ref{eq:xi0_2}), that is in the Kaiser limit,
assuming that the non-linear effects can be neglected at these
scales. In this case, the $b\sigma_8$ factor has to be fixed a
priori. When analysing real catalogues, this factor can be determined
either directly with the deprojection technique \citep[see
  e.g.][]{marulli2012b}, or from the three-point correlation function
\citep[see e.g.][]{moresco2014}, or by combining clustering and
lensing measurements \citep[see e.g.][]{sereno2015}. To minimise the
uncertainties and systematics possibly present in the above methods,
we derive $b\sigma_8$ directly from the real-space two-point
correlation functions of our mock tracers, through
Eq.~(\ref{eq:bias}), while the DM correlation is obtained by Fourier
transforming the linear {\small CAMB} power spectrum, fixing
$\sigma_8$ at the value of the {\em Magneticum} simulation.

The second method considered in this work consists in fully exploiting
the redshift-space two-dimensional anisotropic correlation function
\xiiz. An alternative approach would be to use the multipole moments
of the correlation function. The advantage is to reduce the number of
observables, thus helping in the computation of the covariance matrix
\citep[e.g.][]{delatorre2013b, petracca2016}. However, we are not
mimicking here any real redshift survey. Thus, a detailed treatment of
the statistical errors is not necessary for the purposes of the
present paper, that focuses instead on the systematic uncertainties
caused by a poor modelisation of non-linear effects. Therefore, we
will use just diagonal covariance matrices, whose elements are
estimated analytically from Poisson statistics. As discussed in
\citet{bianchi2012}, the effect on growth rate measurements caused by
assuming negligible off-diagonal elements in the covariance matrix is
small (of the order of only a few percent in that analysis), thanks to
the large volumes of the mock samples with respect to the scales used
for the parameter estimations. Appendix \ref{app:errors} provides some
further tests, using the mock samples analysed in this work, that
appear in overall agreement with the \citet{bianchi2012} conclusions
\citep[see also][]{petracca2016}. Moreover, to investigate the
accuracy of the dispersion model as a function of scales, it is more
convenient to exploit \xiiz. Indeed, following this approach, we can
explore the full $r_\parallel-r_\perp$ plane, searching for the
optimal region to be used to minimise model uncertainties.

\begin{figure}
\includegraphics[width=0.49\textwidth]{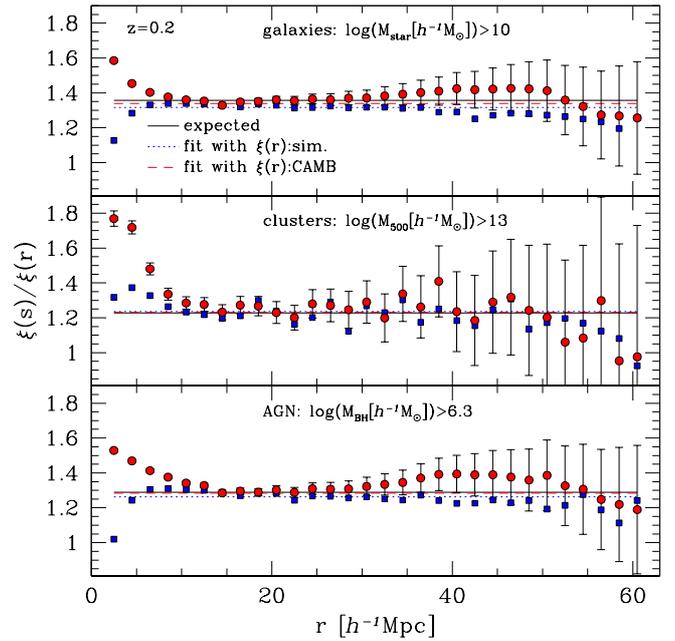}
\caption{The ratio between the redshift-space and real-space two-point
  spherically averaged correlation functions, $\xi(s)/\xi(r)$, at
  $z=0.2$, for galaxies with $\log(M_{\rm STAR} [h^{-1}\,{\mbox
      M_\odot}])>10$ ({\em top panel}), clusters with $\log(M_{500}
  [h^{-1}\,{\mbox M_\odot}])>13$ ({\em central panel}), and AGN with
  $\log(M_{\rm BH} [h^{-1}\,{\mbox M_\odot}])>6.3$ ({\em bottom
    panel}). The redshift-space correlation functions have been
  measured directly from the simulation, while the real-space ones
  have been either measured from the simulation ({\em red dots}), or
  derived from the {\small CAMB} power spectrum ({\em blue
    squares}). The error bars represent the statistical noise as
  prescribed by \citet{mo1992}. The black solid lines show the
  expected values of $\xi(s)/\xi(r)$ at large scales predicted by
  Eq.~(\ref{eq:xi0_2}). The dotted blue and dashed red lines are the
  best-fit ratios estimated from the blue and red points,
  respectively.}
\label{fig:xi_ratio1}
\end{figure}

%%%%%%%%%%%%%%%%%%%%%%%%%%%%%%%%%%%%%%%%%%%%%%%%%%%%%%%%%%%%%%%%%%%%%%%%%%%%%%%

\section{Results}
\label{sec:results}

In this section, we present our main results obtained from mock
samples of three different tracers -- galaxies, clusters and AGN --
using the two approaches described in \S\ref{sec:methodology}. For
each catalogue, we analyse six snapshots corresponding to the
redshifts $z=\{0.2, 0.52, 0.72, 1, 1.5, 2\}$. Moreover, at each
redshift we consider different sample selections. In total, we analyse
$270$ mock catalogues, whose main properties, including the number of
objects in each sample, are reported in Tables \ref{tab:table1},
\ref{tab:table2} and \ref{tab:table3}, and in Fig.~\ref{fig:nz} in
Appendix~\ref{app:samples}.

\begin{figure}
\includegraphics[width=0.49\textwidth]{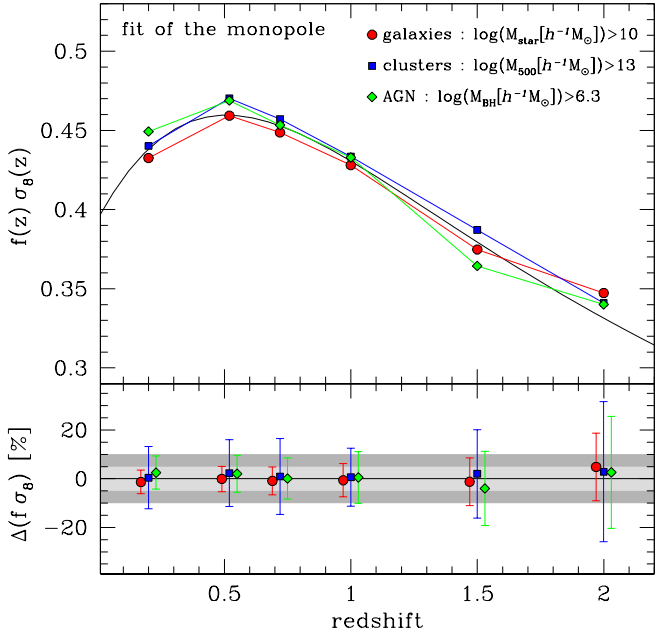}
\caption{{\em Top panel}: the best-fit values of $f(z)\sigma_8(z)$ for
  galaxies with $\log(M_{\rm STAR} [h^{-1}\,{\mbox M_\odot}])>10$
  ({\em red dots}), clusters with $\log(M_{500} [h^{-1}\,{\mbox
      M_\odot}])>13$ ({\em blue squares}), and AGN with $\log(M_{\rm
    BH} [h^{-1}\,{\mbox M_\odot}])>6.3$ ({\em green diamonds}),
  obtained with the method described in \S\ref{subsec:resI}. The black
  line shows the function $\Omega_{\rm M}(z)^{0.545}\cdot\sigma_8(z)$,
  where $\Omega_{\rm M}(z)$ and $\sigma_8(z)$ are the known values of
  the simulation. {\em Bottom panel}: the percentage systematic errors
  on $f(z)\sigma_8(z)$, $[(f\sigma_8)^{\rm
      measured}$-$(f\sigma_8)^{\rm simulation}]/(f\sigma_8)^{\rm
    simulation}\cdot100$, that is the percentage differences between
  the points and the black line shown in the top panel. The error bars
  have been estimated with Eq.~(\ref{eq:bianchi}). The values reported
  have been slightly shifted for visual clarity. To guide the eyes,
  the light and grey shaded areas highlight the $5\%$ and $10\%$ error
  regions, respectively.}
\label{fig:fsigma8}
\end{figure}

%%%%%%%%%%%%%%%%%%%%%%%%%%%%%%%%%%%%%%%%%%%%%%%%%%%%%%%%%%%%%%%%%%%%%%%%%%%%%%%

\subsection{The linear growth rate from the clustering monopole}
\label{subsec:resI}

We start analysing the spherically averaged two-point correlation
function -- the clustering monopole -- at large linear scales. The aim
of this exercise is to investigate the accuracy of the RSD model
in the simplest case possible, that is in the so-called Kaiser limit.

Fig.~\ref{fig:xi_ratio1} shows the ratio between the redshift-space
and real-space two-point correlation functions,
$\xi(s)/\xi(r)$. Specifically, we show here the results obtained at
$z=0.2$, for galaxies with $\log(M_{\rm STAR} [h^{-1}\,{\mbox
    M_\odot}])>10$, clusters with $\log(M_{500} [h^{-1}\,{\mbox
    M_\odot}])>13$, and AGN with $\log(M_{\rm BH} [h^{-1}\,{\mbox
    M_\odot}])>6.3$, as reported by the labels. These correspond to
the upper left samples reported in Tables \ref{tab:table1},
\ref{tab:table2} and \ref{tab:table3}. Results obtained from the other
mock samples are similar, but more scattered due to the lower
densities.  The redshift-space correlation functions have been
measured directly from the simulation, through the procedure described
in \S\ref{sec:methodology}. The real-space correlation functions have
been estimated in two different ways. Either they are measured
directly from the simulation, or they are derived from the linear
{\small CAMB} power spectrum, and assuming a linear bias factor. The
differences at small scales, $r\lesssim5$ \Mpch, between the ratios
computed with the measured (blue squares) and {\small CAMB} (red dots)
real-space correlation functions are due to non-linear effects. We
verified that using the non-linear {\small CAMB} power spectrum, via
the {\small HALOFIT} routine \citep{smith2003}, does not fully remove
the discrepancy. Nevertheless, as we model here the large scale
clustering, this has no effects on our results. On the other hand, the
small discrepancies at scales $r\gtrsim40$ \Mpch can introduce
systematics, that however result smaller than the estimated
uncertainties. The error bars in Fig.~\ref{fig:xi_ratio1} show the
statistical Poisson noise \citep{mo1992}. Scales larger than $60$
\Mpch, not shown in the plot, are too noisy to affect the fit. More
specifically, we verified that a convenient scale range to get robust
results is $10<r[h^{-1}\,\mbox{Mpc}]<50$. The black solid lines show
the expected values of $\xi(s)/\xi(r)$ at large scales, as predicted
by Eq.~(\ref{eq:xi0_2}).

By modelling the clustering ratio, $\xi(s)/\xi(r)$, via
Eq.~(\ref{eq:xi0_1}), it is possible to estimate the factor
$\beta$. We do this by measuring both $\xi(s)$ and $\xi(r)$ from the
simulation. The result is shown by the dotted blue lines in
Fig.~\ref{fig:xi_ratio1}. Instead, to estimate $f\sigma_8$ from the
monopole of the two-point correlation function, we use
Eq.~(\ref{eq:xi0_2}). In this case the factor $b\sigma_8$ has to be
fixed. This requires to estimate the real-space two-point correlation
function of the DM. As described above, we obtain the latter by
Fourier transforming the {\small CAMB} power spectrum. Moreover, to
estimate the mean linear bias, a range of scales has to be chosen in
Eq.~(\ref{eq:bias}). We compute the linear bias in the range
$\bar{r}<r[$\Mpch$]<50$, where $\bar{r}$ is a free parameter that sets
the minimum scale beyond which the bias is fairly
scale-independent. Such a minimum scale depends on redshift and sample
selection, and we estimate it for each sample considered. The dashed
red lines are the best-fit ratios estimated with this method.

\begin{figure}
  \includegraphics[width=0.48\textwidth]{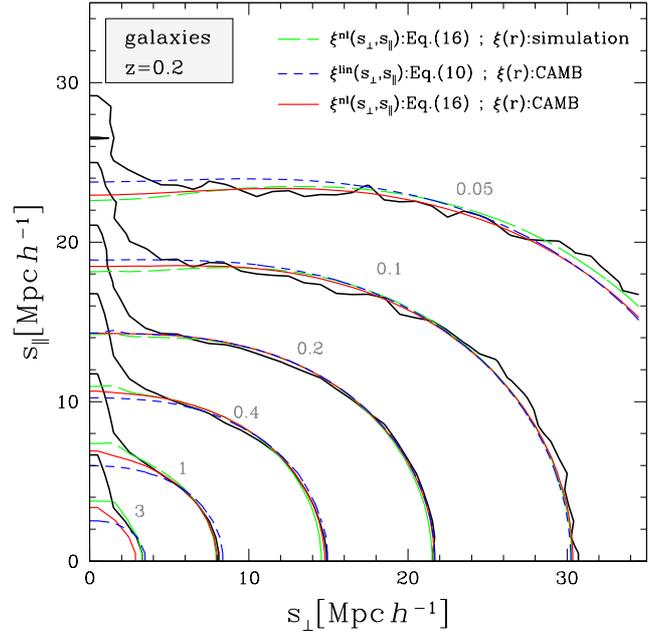}
  \caption{The iso-correlation contours of the redshift-space
    two-point correlation function, corresponding to the values
    \xiiz$=[0.05, 0.1, 0.2, 0.4, 1, 3]$, for galaxies with
    $\log(M_{\rm STAR} [h^{-1}\,{\mbox M_\odot}])>10$, at $z=0.2$
    (black contours). The dot-dashed green and solid red contours show
    the best-fit model given by Eq.~(\ref{eq:ximodel}), with the
    real-space correlation function $\xi(r)$ measured from the
    simulation, and estimated from the {\small CAMB} power spectrum,
    respectively. The blue dashed contours show the linear best-fit
    model given by Eq.~(\ref{eq:ximodellin}) with the {\small CAMB}
    real-space correlation function.}
  \label{fig:iso_gal}
\end{figure}

\begin{figure}
  \includegraphics[width=0.48\textwidth]{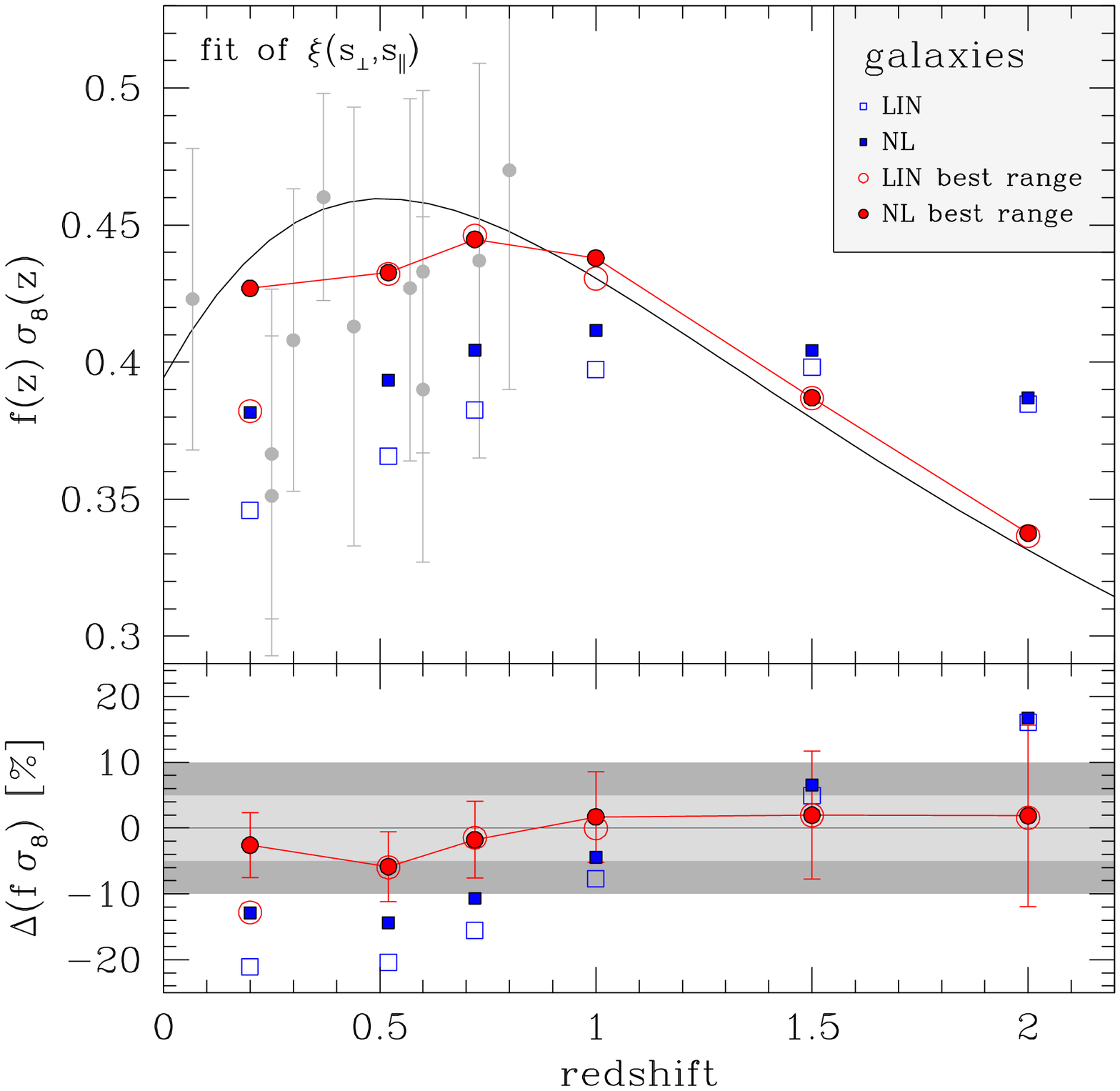}
  \includegraphics[width=0.48\textwidth]{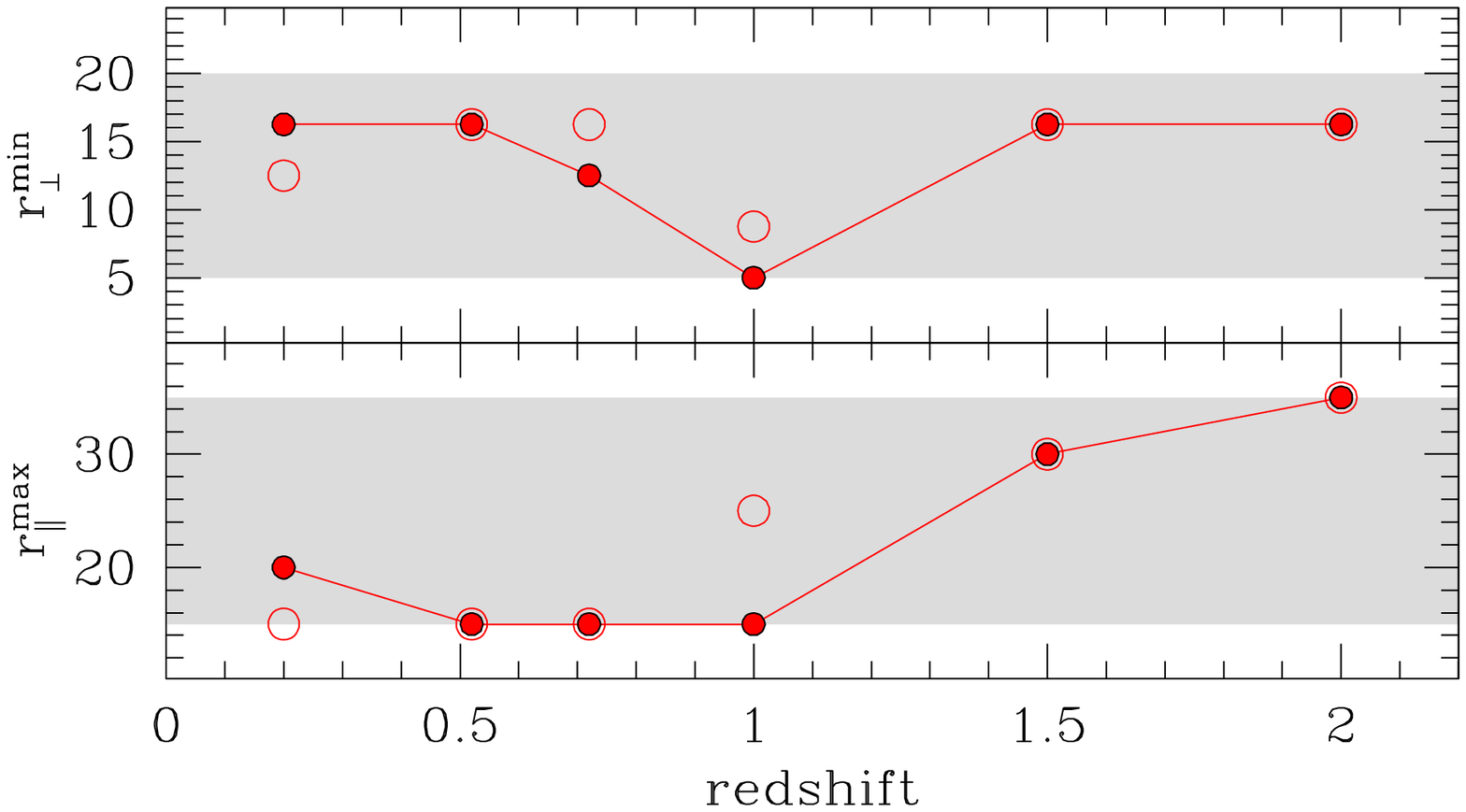}
  \caption{{\em Top panel of the upper window}: the best-fit values of
    $f(z)\sigma_8(z)$ for galaxies with $\log(M_{\rm STAR}
    [h^{-1}\,{\mbox M_\odot}])>10$. The open and solid blue squares
    show the values obtained by fitting the data with the models given
    by Eqs.~(\ref{eq:ximodellin}) and (\ref{eq:ximodel}),
    respectively, in the scale ranges
    $3<r_\perp[h^{-1}\,\mbox{Mpc}]<35$ and
    $3<r_\parallel[h^{-1}\,\mbox{Mpc}]<35$. The open and solid red
    dots have been obtained with the method described in
    \S\ref{subsubsec:galaxies}, fitting the data with the models given
    by Eqs.~(\ref{eq:ximodellin}) and (\ref{eq:ximodel}),
    respectively. The black line shows the function $\Omega_{\rm
      M}(z)^{0.545}\cdot\sigma_8(z)$, where $\Omega_{\rm M}(z)$ and
    $\sigma_8(z)$ are the known values of the simulation. For
    comparison, the grey dots show a set of recent observational
    measurements from galaxy surveys (see
    \S\ref{subsubsec:galaxies}). {\em Bottom panel of the upper
      window}: the percentage systematic errors on $f(z)\sigma_8(z)$,
    defined as $[(f\sigma_8)^{\rm measured}$-$(f\sigma_8)^{\rm
        simulation}]/(f\sigma_8)^{\rm simulation}\cdot100$, that is
    the percentage differences between the points and the black line
    shown in the top panel. To guide the eyes, the light and grey
    shaded areas highlight the $5\%$ and $10\%$ error regions. {\em
      Lower window}: best-fit values of $r_\perp^{\rm min}$ and
    $r_\parallel^{\rm max}$ of the method described in
    \S\ref{subsubsec:galaxies}.  The grey areas show the regions
    explored by the method.  }
  \label{fig:fs8_gal}
\end{figure}

As it can be seen by comparing red and blue lines, the two methods
provide concordant results, both of them in agreement with the
expectations. Indeed, the real-space correlation function provided by
{\small CAMB} and assuming a linear bias is in reasonable agreement
with the one measured directly from the simulation.

The constraints on $f\sigma_8$ as a function of redshift are shown in
the upper panel of Fig.~\ref{fig:fsigma8}. As in
Fig.~\ref{fig:xi_ratio1}, we show the results obtained for galaxies
with $\log(M_{\rm STAR} [h^{-1}\,{\mbox M_\odot}])>10$, clusters with
$\log(M_{500} [h^{-1}\,{\mbox M_\odot}])>13$, and AGN with
$\log(M_{\rm BH} [h^{-1}\,{\mbox M_\odot}])>6.3$. The black line shows
the function $\Omega_{\rm M}(z)^{0.545}\cdot\sigma_8(z)$, where
$\Omega_{\rm M}(z)$ and $\sigma_8(z)$ are the values of the {\em
  Magneticum} simulation.  In the lower panel we show the percentage
systematic errors on $f\sigma_8$, defined as $[(f\sigma_8)^{\rm
    measured}$-$(f\sigma_8)^{\rm simulation}]/(f\sigma_8)^{\rm
  simulation}\cdot100$. The error bars have been estimated by
propagating on $f\sigma_8$ the $\beta$ errors provided by the scaling
formula presented in \citet{bianchi2012}, that gives the statistical
errors as a function of bias, $b$, volume, $V$, and density, $n$:
\begin{equation}
\frac{\delta\beta}{\beta}\simeq
Cb^{0.7}V^{-0.5}\exp\left(\frac{n_0}{b^2n}\right) \; ,
\label{eq:bianchi} 
\end{equation}
where $n_0=1.7\cdot10^{-4}\,h^{3}\,\mbox{Mpc}^{-3}$ and
$C=4.9\cdot10^2 \,h^{-1.5}\,\mbox{Mpc}^{1.5}$. In this case, these
error bars have to be considered just as lower limits, as the scaling
formula has been calibrated based on fits of the full two-dimensional
anisotropic correlation function \xiiz. Moreover, they have been
computed in different scale ranges and with Friends-of-Friends DM
haloes, differently from this analysis, where we consider tracers
hosted in DM sub-haloes. Using the full covariance matrix in this
statistical analysis does not change our conclusions, as shown in
Appendix \ref{app:errors}.

\begin{figure}
  \includegraphics[width=0.48\textwidth]{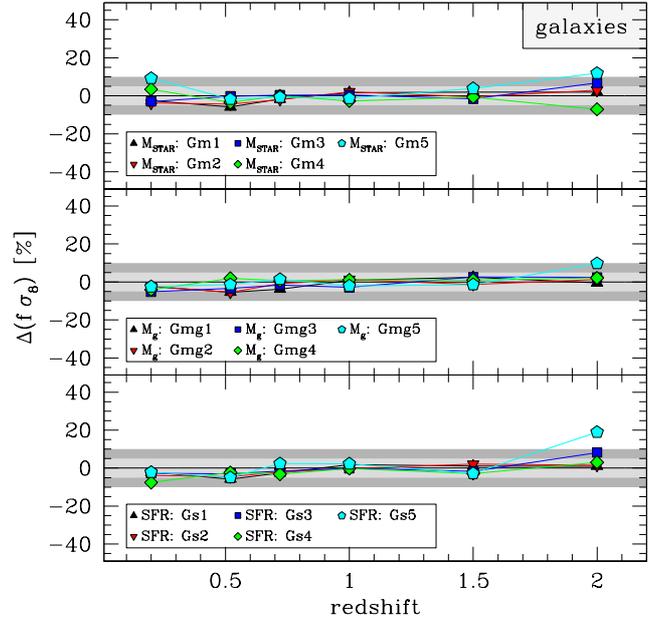}
  \caption{The percentage systematic errors on $f(z)\sigma_8(z)$ for
    galaxies, $[(f\sigma_8)^{\rm measured}$-$(f\sigma_8)^{\rm
        simulation}]/(f\sigma_8)^{\rm simulation}\cdot100$, as a
    function of redshift and sample selections, as indicated by the
    labels. $S1-S5$ refer to the five selection thresholds reported in
    Tables \ref{tab:table1}, \ref{tab:table2} and \ref{tab:table3},
    for the different tracers and properties used for the selection.}
  \label{fig:errfs8_gal}
\end{figure}

As demonstrated by Figs.~\ref{fig:xi_ratio1} and \ref{fig:fsigma8}, it
is indeed possible to get almost unbiased constraints on $f\sigma_8$
from the monopole of the two-point correlation function, at all
redshifts considered, and for both galaxies, clusters and AGN,
provided that the linear bias is estimated at sufficiently large
scales. The advantage of this method is the minimal number of free
parameters necessary for the modelisation. However, the linear bias
has to be assumed, or measured from other probes.

%%%%%%%%%%%%%%%%%%%%%%%%%%%%%%%%%%%%%%%%%%%%%%%%%%%%%%%%%%%%%%%%%%%%%%%%%%%%%%%

\subsection{The linear growth rate from the anisotropic 2D clustering}
\label{subsec:resII}

To extract the full information from the redshift-space two-point
correlation function, the anisotropic correlation \xiiz or,
alternatively, all the relevant multipoles have to be modelled. As
discussed in \S\ref{sec:methodology}, we consider the first
approach. Differently from the analysis of \S\ref{subsec:resI}, here
we can jointly constrain the two terms $f\sigma_8$ and $b\sigma_8$. In
this section we investigate the accuracy of the constraints on
$f\sigma_8$ provided by the dispersion model, as a function of sample
selection, with $b\sigma_8$ and $\sigma_{12}$ as free parameters.

%%%%%%%%%%%%%%%%%%%%%%%%%%%%%%%%%%%%%%%%%%%%%%%%%%%%%%%%%%%%%%%%%%%%%%%%%%%%%%%

\subsubsection{Galaxies}
\label{subsubsec:galaxies}

We start presenting our results for the galaxy mock samples. The black
lines of Fig.~\ref{fig:iso_gal} show the iso-correlation contours of
the redshift-space two-point correlation function, \xiiz, for galaxies
with $\log(M_{\rm STAR} [h^{-1}\,{\mbox M_\odot}])>10$, at
$z=0.2$. The other lines are the best-fit models obtained with three
different methods. The dot-dashed green lines are obtained by using
the full non-linear dispersion model given by Eq.~(\ref{eq:ximodel}),
with the real-space correlation function $\xi(r)$ measured directly
from the simulation. The blue dashed and red solid contours are
obtained with $\xi(r)$ estimated from the {\small CAMB} power
spectrum, and by fitting the linear model given by
Eq.~(\ref{eq:ximodellin}) and the non-linear one by
Eq.~(\ref{eq:ximodel}), respectively. The fitting is done in the scale
ranges $3<r_\perp[h^{-1}\,\mbox{Mpc}]<35$ and
$3<r_\parallel[h^{-1}\,\mbox{Mpc}]<35$.

As one can see, while all the three models provide a good description
of the anisotropic clustering at scales larger than $\sim5$ \Mpch,
none of them can reproduce accurately the fingers-of-God pattern at
small scales. This is expected for the blue contours, obtained without
modelling the non-linear dynamics. For the other two cases, the
discrepancy is due to the not sufficiently accurate description of
non-linearities provided by the dispersion model. We verified that the
minimum scales considered for the fit have a negligible impact on the
fingers-of-God shape, that is we find the same results even
considering scales smaller that 3 \Mpch in the fit. The green contours
appear in better agreement with the measurements at small scales, due
to the fact that the real-space correlation function is measured from
the mock catalogue\footnote{The model corresponding to the green
  contours provides constraints on $\beta$, not on $f\sigma_8$. We
  show it here only to highlight the differences at small scales
  between the dispersion model estimated with $\xi(r)$ measured from
  the simulation, and the one with $\xi(r)$ from the linear {\small
    CAMB} power spectrum, assuming a linear scale-independent bias.}.

\begin{figure}
  \includegraphics[width=0.48\textwidth]{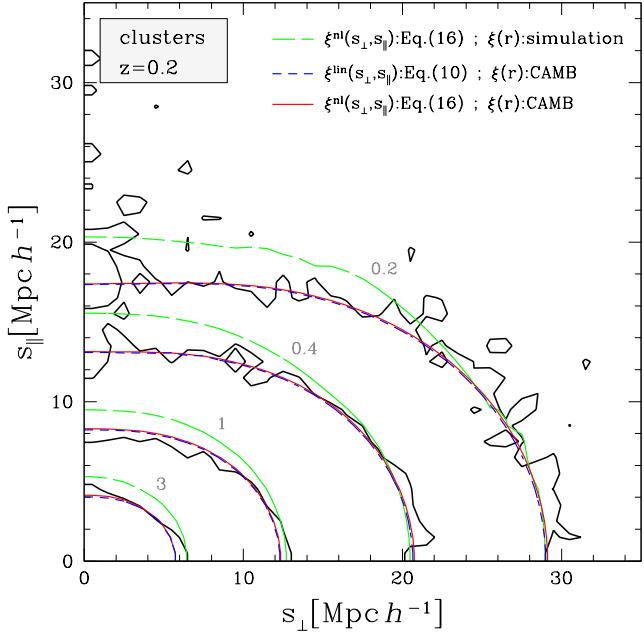}
  \caption{The iso-correlation contours of the redshift-space
    two-point correlation function, corresponding to the values
    \xiiz$=[0.2, 0.4, 1, 3]$, for clusters with $\log(M_{500}
         [h^{-1}\,{\mbox M_\odot}])>13$, at $z=0.2$ (black
         contours). The other contours are as in
         Fig.~\ref{fig:iso_gal}.}
  \label{fig:iso_cl}
\end{figure}

\begin{figure}
  \includegraphics[width=0.48\textwidth]{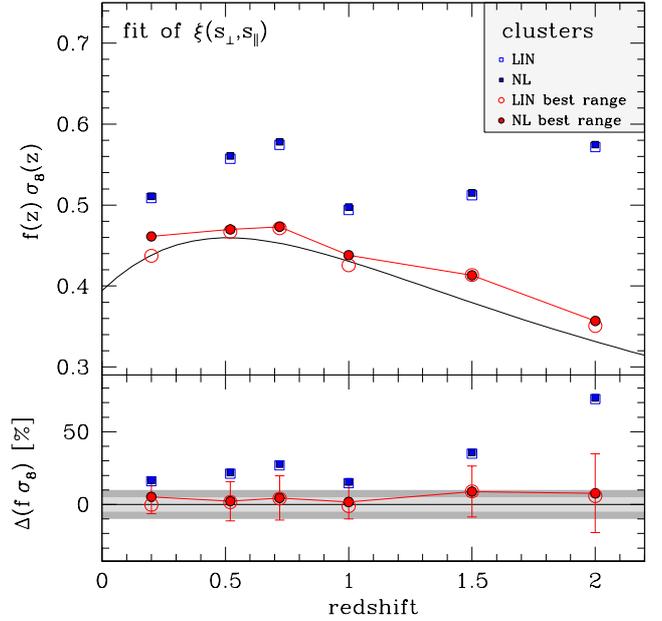}
  \includegraphics[width=0.48\textwidth]{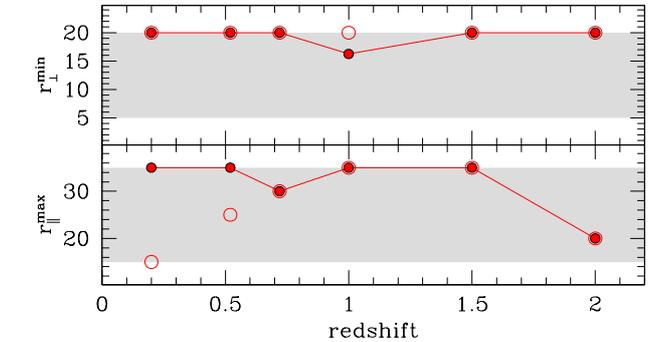}
  \caption{The best-fit values of $f(z)\sigma_8(z)$ for clusters with
    $\log(M_{500} [h^{-1}\,{\mbox M_\odot}])>13$. All the symbols are
    as in Fig.~\ref{fig:fs8_gal}.}
  \label{fig:fs8_cl}
\end{figure}

\begin{figure}  
  \includegraphics[width=0.48\textwidth]{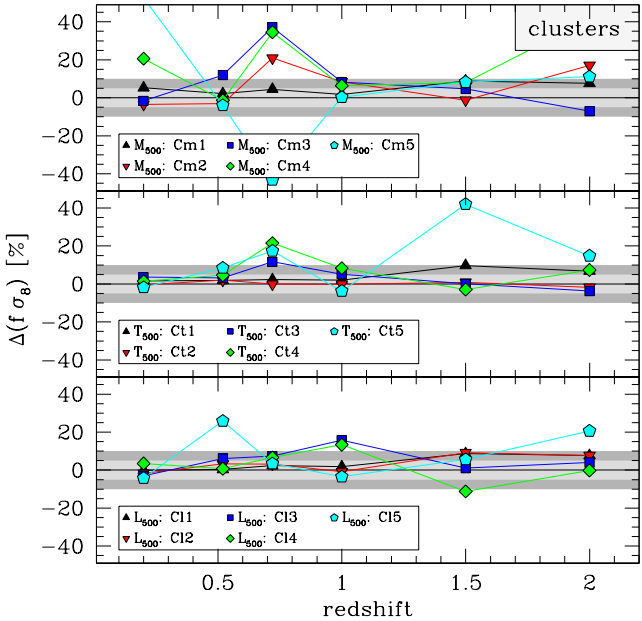}
  \caption{The percentage systematic errors on $f(z)\sigma_8(z)$ for
    clusters, as a function of redshift and sample selections. All
    symbols are as in Fig.~\ref{fig:errfs8_gal}.}
  \label{fig:errfs8_cl}
\end{figure}

The best-fit values of $f\sigma_8$ as a function of redshift are shown
in the top panel of the upper window of Fig.~\ref{fig:fs8_gal}. The
open and solid blue squares show the values obtained by fitting the
data with the models given by Eqs.~(\ref{eq:ximodellin}) and
(\ref{eq:ximodel}), respectively, that is they correspond to the blue
and red contours of Fig.~\ref{fig:iso_gal} (at $z=0.2$). The bottom
panel of the upper window shows the percentage systematic errors on
$f\sigma_8$, that is $[(f\sigma_8)^{\rm measured}$-$(f\sigma_8)^{\rm
    simulation}]/(f\sigma_8)^{\rm simulation}\cdot100$.

As it can be seen, the best-fit values of $f\sigma_8$ result strongly
biased with respect to the true ones, with a systematic error that
depends on the redshift. At $z=0.2$, we get an underestimation of
$\sim10$ ($20$) $\%$ with the non-linear (linear) dispersion model, at
$z=1$ the error reduces to $\sim5$ ($8$)$\%$, while at $z=2$ we get an
overestimation of $\sim20\%$, with both linear and non-linear
modelling. This result is fairly in agreement with what found by
\citet{bianchi2012} at $z=1$, although the two analyses are not
directly comparable, as \citet{bianchi2012} considered a sample of
Friends-of-Friends DM haloes, slightly affected by fingers-of-God. As
expected, the convolution given by Eq.~(\ref{eq:ximodel}) reduces the
systematic error on $f\sigma_8$, especially at low redshifts. However,
it does not totally remove the discrepancies, in agreement with
previous findings \citep[e.g.][and references
  therein]{mohammad2016}. For comparison, the grey dots of
Fig.~\ref{fig:fs8_gal} show a set of recent observational measurements
of $f\sigma_8$ from large galaxy surveys: 6dFGS at $z=0.067$
\citep{beutler2012}; SDSS(DR7) Luminous Red Galaxies at $z=0.25, 0.37$
\citep{samushia2012} from scales lower than $60$ and $200$ \Mpch; BOSS
at $z=0.3, 0.57, 0.6$ \citep{tojeiro2012, reid2012}; WiggleZ at
$z=0.44, 0.6, 0.73$ \citep{blake2012}; VIPERS at $z=0.8$
\citep{delatorre2013b}. These results have been obtained using
different RSD models, most of them more accurate than the dispersion
model considered in this work. Nevertheless, many of these
measurements underestimate $f\sigma_8$ with respect to GR+$\Lambda$CDM
predictions \citep[see e.g.][]{macaulay2013}.  In line with our
findings, this could be explained, at least partially, by model
uncertainties still present in the more sophisticated RSD models
considered.

The direct way to reduce these systematics is to improve the
modelisation of RSD at non-linear scales \citep[see
  e.g.][]{scoccimarro2004, taruya2010, seljak2011, wang2014,
  delatorre2013b}. We explore here a different approach, investigating
the dependency of the systematic error on the comoving scales
considered in the analysis. In the forthcoming plots, we show the
results obtained by repeating our fitting procedure for different
values of the minimum perpendicular separation and the maximum
parallel separation used in the fit, that is $r_\perp^{\rm min}$ and
$r_\parallel^{\rm max}$. By increasing the value of $r_\perp^{\rm
  min}$ we can cut the region more affected by fingers-of-God
anisotropies, while by reducing $r_\parallel^{\rm max}$ we avoid the
region more affected by shot noise. As we verified, changing also
$r_\perp^{\rm max}$ and $r_\parallel^{\rm min}$, or adopting different
scale selection criteria, do not affect significantly the results.
The aim here is to search for optimal regions in this plane to
possibly get unbiased constraints. As non-linear dynamics impact on
different scales for different redshifts and biases of the tracers, we
expect that the best values of $r_\perp^{\rm min}$ and
$r_\parallel^{\rm max}$, that is the ones that minimise systematics,
will be different for different sample selections.

The results of this analysis are shown in Fig.~\ref{fig:fs8_gal} with
open and solid red dots, obtained by fitting the data with the model
given by Eqs.~(\ref{eq:ximodellin}) and (\ref{eq:ximodel}),
respectively, that is by considering either the linear or the
non-linear RSD model.  We explore the ranges $5<r_\perp^{\rm
  min}[h^{-1}\,\mbox{Mpc}]<20$ and $15<r_\parallel^{\rm
  max}[h^{-1}\,\mbox{Mpc}]<35$, highlighted by the grey areas in the
bottom panels of the lower window of Fig.~\ref{fig:fs8_gal}.
Specifically, we consider a 10x10 grid in the $r_\perp^{\rm min}$,
$r_\parallel^{\rm max}$ plane. The best-fit values of $r_\perp^{\rm
  min}$ and $r_\parallel^{\rm max}$, shown in the bottom panels of the
lower window, are the ones that minimise the systematic error on
$f\sigma_8$. The ranges considered are large enough to get systematic
errors on $f\sigma_8$ lower than $\sim 5\%$. Indeed, for proper values
of $r_\perp^{\rm min}$ and $r_\parallel^{\rm max}$, it is possible to
significantly reduce the systematic error on $f\sigma_8$ at all
redshifts, without having to improve the treatment of
non-linearities. The error bars have been estimated using the scaling
formula given by Eq.~\ref{eq:bianchi}.

Both the linear and non-linear dispersion models provide similar
results when $r_\perp^{\rm min}$ and $r_\parallel^{\rm max}$ are kept
free. In other words, it seems more convenient simply to not consider
the scales where non-linear effects have a non-negligible impact,
rather than to try modelling them with the empirical description
provided by Eq.~(\ref{eq:ximodel}).

These results show that i) it is possible to reduce significantly the
systematics in RSD constraints, even with the dispersion model, by
cutting the $r_\perp^{\rm min}$, $r_\parallel^{\rm max}$ plane
conveniently, and ii) the {\em optimal} $r_\perp^{\rm min}$,
$r_\parallel^{\rm max}$ range depends on sample selection, thus it
cannot be just fixed in multi-tracer analyses. As it can be seen in
Fig.~\ref{fig:fs8_gal}, we do not find any clear trend of
$r_\perp^{\rm min}$ and $r_\parallel^{\rm max}$ as a function of
redshift, as it would be expected if the systematics were caused by
non-linearities. This is possibly caused by the not sufficiently large
volumes considered. The analysis of larger data sets, that we defer to
a future work, will hopefully clarify this point.

\begin{figure}
  \includegraphics[width=0.48\textwidth]{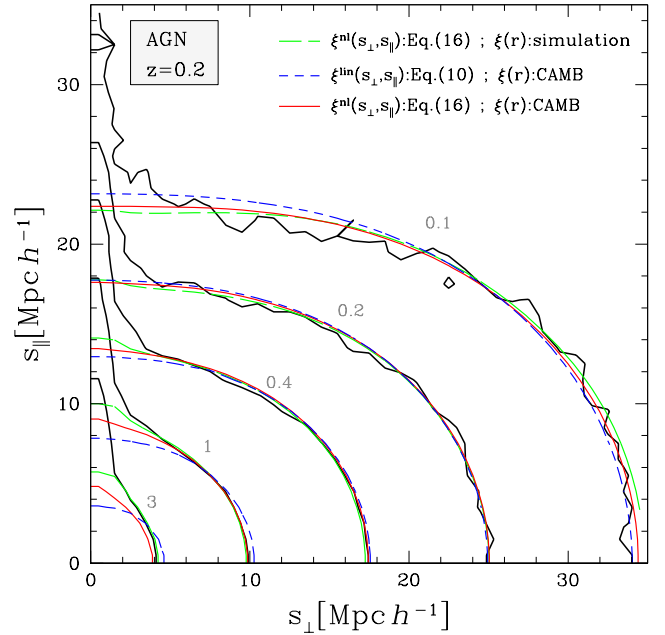}
  \caption{The iso-correlation contours of the redshift-space
    two-point correlation function, corresponding to the values
    \xiiz$=[0.1, 0.2, 0.4, 1, 3]$, for AGN with $\log(M_{\rm BH}
         [h^{-1}\,{\mbox M_\odot}])>6.3$, at $z=0.2$ (black
         contours). The other contours show the best-fit models, as
         in Fig.~\ref{fig:iso_gal}.}
  \label{fig:iso_agn}
\end{figure}

\begin{figure}
  \includegraphics[width=0.48\textwidth]{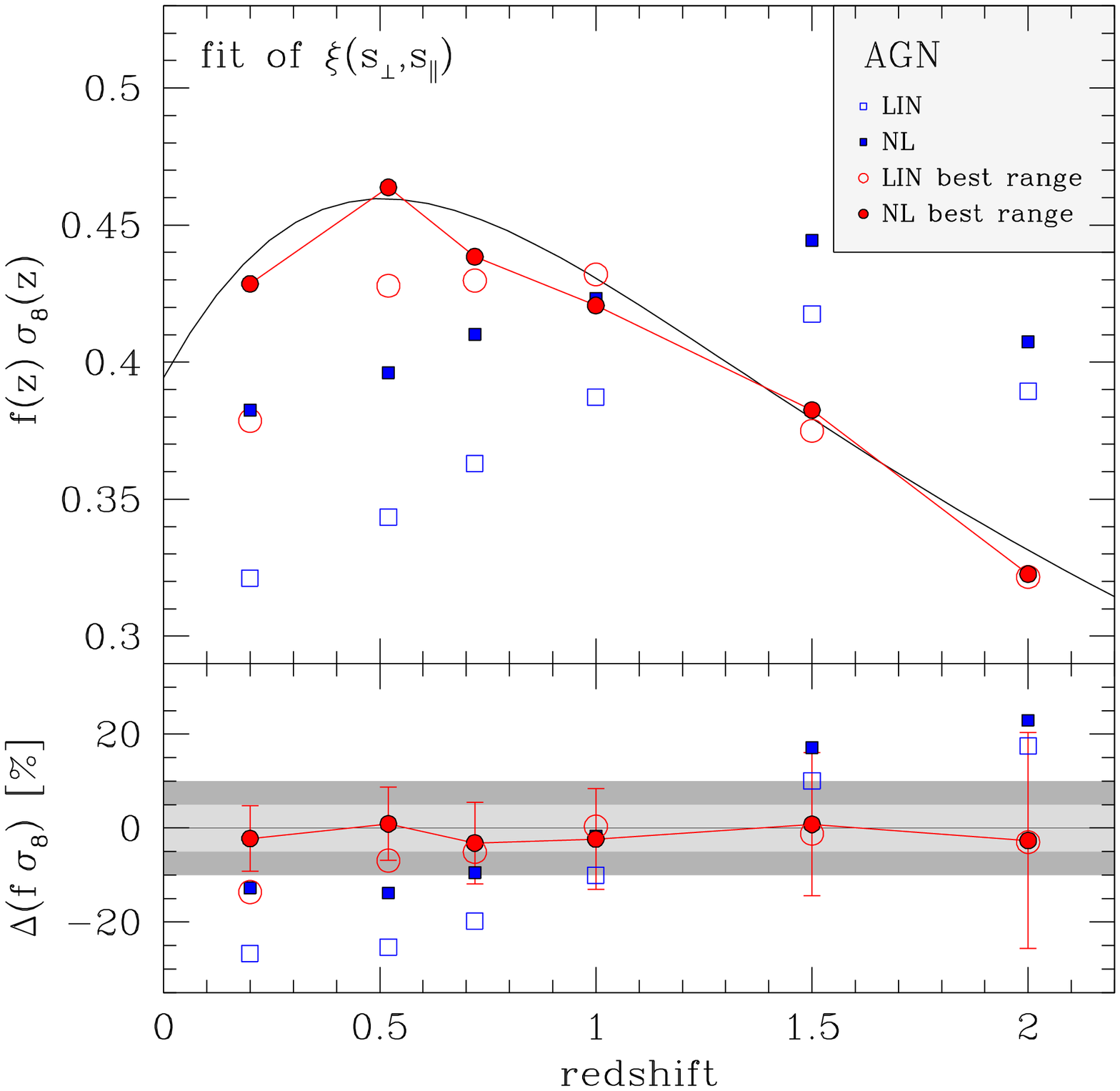}
  \includegraphics[width=0.48\textwidth]{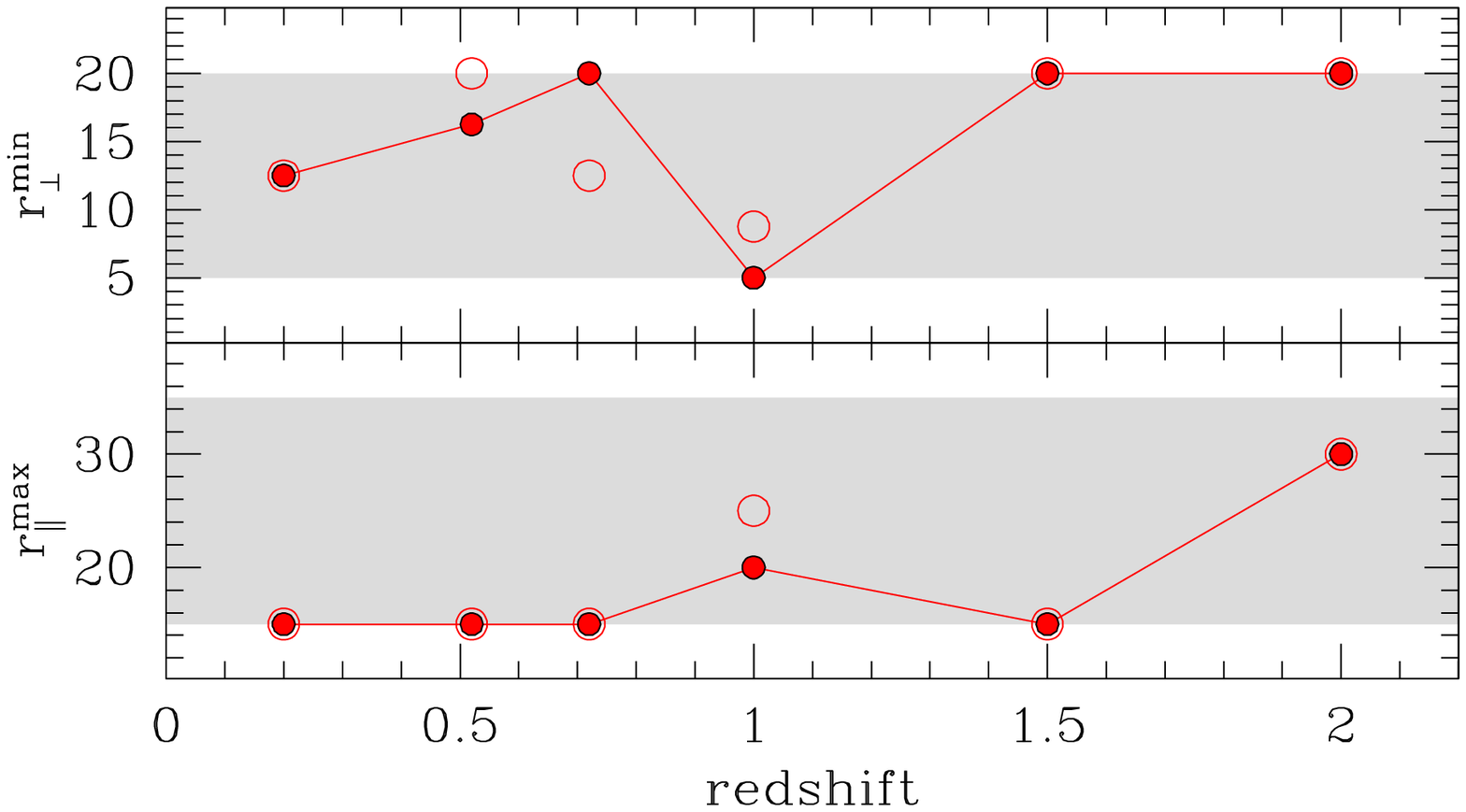}
  \caption{The best-fit values of $f(z)\sigma_8(z)$ for AGN with
    $\log(M_{\rm BH} [h^{-1}\,{\mbox M_\odot}])>6.3$. All the symbols
    are as in Fig.~\ref{fig:fs8_gal}.}
  \label{fig:fs8_agn}
\end{figure}

\begin{figure}  
  \includegraphics[width=0.48\textwidth]{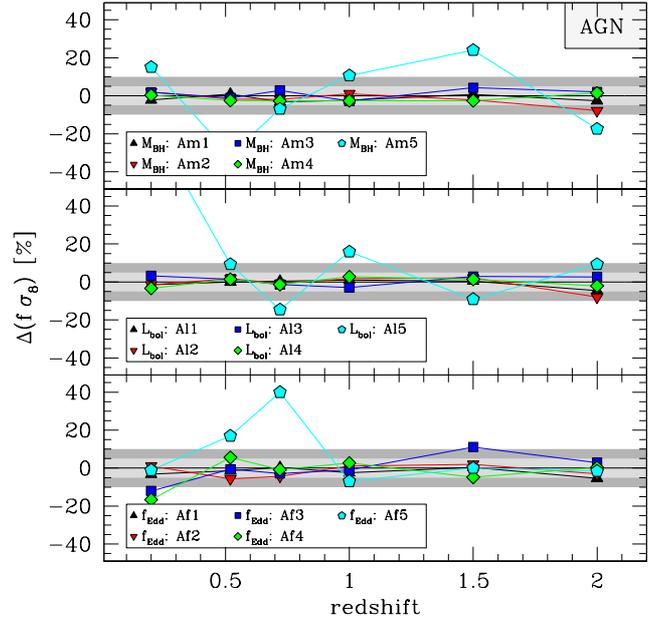}
  \caption{The percentage systematic errors on $f(z)\sigma_8(z)$ for
    AGN, as a function of redshift and sample selections. All symbols
    are as in Fig.~\ref{fig:errfs8_gal}.}
  \label{fig:errfs8_agn}
\end{figure}

For completeness, we then apply our method to several subsamples with
different selections. Specifically, we consider galaxy catalogues
selected in stellar mass, $M_{\rm STAR}$, g-band absolute magnitude,
$M_g$, and star formation rate, SFR. The selection thresholds
considered and the number of galaxies in each sample are reported in
Table \ref{tab:table1}. The percentage systematic errors on
$f\sigma_8$ are reported in Fig.~\ref{fig:errfs8_gal}, as a function
of redshift and sample selections, as indicated by the labels. The
result is quite remarkable: it is indeed possible to get almost
unbiased constraints on $f\sigma_8$ with the dispersion model,
independent of sample selections, provided that $r_\perp^{\rm min}$
and $r_\parallel^{\rm max}$ are chosen conveniently. For the largest
galaxy sample shown in Fig.~\ref{fig:fsigma8}, the preferred value of
$r_\perp^{\rm min}$ is $\sim 15$ \Mpch at low and high redshifts, and
$\sim 5$ \Mpch around $z=1$, while $r_\parallel^{\rm max}$ is $\sim
15$ \Mpch at low redshifts, and increases up to $\sim 35$ \Mpch at
$z=2$. Again, we do not find any clear trend of $r_\perp^{\rm min}$
and $r_\parallel^{\rm max}$ as a function of redshift or sample bias.

Overall, all of these results show that our modelling of RSD is mostly
sensitive to the lower fitting cut-off, depending on how the tracers
relate to the underlying mass. Indeed, the effect of the different
selections considered here is just to select subsamples of haloes of
different masses (hence bias) hosting the observed galaxies. As noted
above, this has to be considered in multi-tracer analyses, where a
single sample selection might introduce systematics.

%%%%%%%%%%%%%%%%%%%%%%%%%%%%%%%%%%%%%%%%%%%%%%%%%%%%%%%%%%%%%%%%%%%%%%%%%%%%%%%

\subsubsection{Galaxy groups and clusters}
\label{subsubsec:clusters}

In this section, we perform the same analysis presented in
\S\ref{subsubsec:galaxies} on the mock samples of galaxy groups and
clusters. To simplify the discussion, in the following we will use the
term {\em clusters} to refer to all of these objects, though the less
massive ones should be considered as galaxy groups, or just haloes,
from an observational perspective (see Table \ref{tab:table2}).

Fig.~\ref{fig:iso_cl} shows the iso-correlation contours of the
redshift-space two-point correlation function of clusters with
$\log(M_{500} [h^{-1}\,{\mbox M_\odot}])>13$, at $z=0.2$. The other
contours are the best-fit models, as in Fig.~\ref{fig:iso_gal}. As it
can be seen, the fingers-of-God anisotropies are almost absent,
differently from the galaxy correlation function shown in
Fig.~\ref{fig:iso_gal}, due to the lower values of non-linear motions
of clusters at small scales. This makes the convolution of
Eq.~(\ref{eq:ximodel}) negligible, as it can be seen comparing the
dashed blue and solid red lines. Interestingly, when the real-space
correlation function $\xi(r)$ is measured from the mocks (green
contours), the dispersion model fails to match the small-scale
clustering shape. This is primarily caused by the paucity of the
cluster sample. Due to the low number of cluster pairs at small
separations, the small-scale clustering cannot be estimated
accurately, thus introducing systematics in the model. On the
contrary, when $\xi(r)$ is estimated from {\small CAMB}, the
small-scale clustering shape can be accurately described. For visual
clarity, we have shown here only the iso-correlation contours down to
\xiiz$=0.2$, as for lower values they appear too scattered.

Fig.~\ref{fig:fs8_cl} shows the best-fit values of $f\sigma_8$ as a
function of redshift. In contrast with what we found with galaxies,
the constraints on the linear growth rate obtained in the scale ranges
$3<r_\perp[h^{-1}\,\mbox{Mpc}]<35$ and
$3<r_\parallel[h^{-1}\,\mbox{Mpc}]<35$ (blue squares) result severely
overestimated at all redshifts, independent of using the linear or the
non-linear dispersion model. To investigate how these systematics
depend on the scale range considered in the fit, we repeat the same
analysis described in \S\ref{subsubsec:galaxies}, keeping free the
parameters $r_\perp^{\rm min}$ and $r_\parallel^{\rm max}$. The result
is shown by the red dots in Fig.~\ref{fig:fs8_cl}. As one can see, we
are able to substantially reduce the systematics on $f\sigma_8$, at
all redshifts. This is obtained by cutting the small perpendicular
separations, limiting the analysis at $r_\perp\gtrsim20$ \Mpch, as
shown in the lower window of Fig.~\ref{fig:fs8_cl}.

These results appear in overall agreement with what previously found
by \citet{bianchi2012}, where systematic errors become positive for DM
halo masses larger than $10^{13}h^{-1}\,{\mbox M_\odot}$, that is for
group masses and above. This represents a nice confirmation, not
related to the specific simulation or halo/cluster selection, but only
to the dynamics of haloes of large masses.

Then, we consider several cluster subsamples selected in mass,
$M_{500}$, temperature, $T_{500}$, and luminosity, $L_{500}$, whose
main properties are reported in Table
\ref{tab:table2}\footnote{$M_{500}$, $T_{500}$ and $L_{500}$ are,
  respectively, the mass, the temperature and the luminosity enclosed
  within a sphere of radius $r_{500c}$, in which the mean matter
  density is equal to $500$ times the critical density $\rho_{\rm
    crit}(z) = 3H^2(z)/8\pi G$, where $H(z)$ is the Hubble
  parameter.}.  Fig.~\ref{fig:errfs8_cl} shows the percentage
systematic errors on $f\sigma_8$ as a function of redshift and sample
selections. For the largest samples analysed we do not get any
significant bias on $f\sigma_8$, similarly to what we found with
galaxies. However, when selecting too few objects, the results appear
quite scattered, though without systematic trends. Indeed, the
advantages resulting from the low fingers-of-God anisotropies
compensate with the paucity of cluster pairs at small
scales. Summarising, it seems possible to get unbiased constraints on
the linear growth rate from the RSD of galaxy clusters, but the
cluster sample has to be sufficiently numerous.

%%%%%%%%%%%%%%%%%%%%%%%%%%%%%%%%%%%%%%%%%%%%%%%%%%%%%%%%%%%%%%%%%%%%%%%%%%%%%%%

\subsubsection{AGN}
\label{subsubsec:AGN}

Finally, we analyse the AGN mock samples.  The results are shown in
Figs.~\ref{fig:iso_agn}, \ref{fig:fs8_agn} and \ref{fig:errfs8_agn},
that are the analogous of the plots presented in the previous two
sections for galaxies and clusters. As expected, we find similar
results to the ones found with galaxies. Differently from the cluster
case, the fingers-of-God are clearly evident in the AGN correlation
function, and not accurately described by the dispersion model, as it
can be seen in Fig.~\ref{fig:iso_agn}. Nevertheless, the convolution
of Eq.~(\ref{eq:ximodel}) significantly helps in reducing the
systematic error on $f\sigma_8$, even more than what we found when
analysing the galaxy sample, as evident by comparing
Figs.~\ref{fig:iso_gal} and \ref{fig:iso_agn}. Also in this case, it
is possible to get almost unbiased constraints on $f\sigma_8$ at all
redshifts and for the different selections considered, provided that
the analysis is restricted at $r_\perp^{\rm min}\gtrsim15$ \Mpch and
$r_\parallel^{\rm max}\lesssim15$ \Mpch. Specifically, we considered
here AGN catalogues selected in BH mass, $M_{\rm BH}$, bolometric
luminosity, $L_{\rm bol}$, and Eddington factor, $f_{\rm Edd}$, as
reported in Table \ref{tab:table3}.  Only the AGN samples with the
lower number of objects analysed here (cyan points in
Fig.~\ref{fig:errfs8_agn}) are not large enough to provide robust
constraints on $f\sigma_8$.

%%%%%%%%%%%%%%%%%%%%%%%%%%%%%%%%%%%%%%%%%%%%%%%%%%%%%%%%%%%%%%%%%%%%%%%%%%%%%%%

\section{Discussion}
\label{sec:discussion}
The results presented in this paper extend the previous work by
\citet{bianchi2012}, who analysed mock samples of DM haloes at
$z=1$. One of the main differences with respect to that work is that
here we used galaxy, cluster and AGN simulated catalogues, that are
more directly comparable to observational samples. Moreover, we
investigated the systematic errors of $f\sigma_8$, instead of $\beta$,
and extended the analysis at different redshifts and sample
selections. The use of these simulated objects as {\em test particles}
provides a better treatment of non-linearities in small-scale
dynamics, that translates to a more realistic description of the
fingers-of-God in the redshift-space galaxy and AGN clustering, as it
can be seen in Figs.~\ref{fig:iso_gal} and \ref{fig:iso_agn}. On the
other hand, the selected cluster samples are more similar to the DM
halo samples analysed by \citet{bianchi2012}, as expected, with
fingers-of-God anisotropies almost absent, as shown in
Fig.~\ref{fig:iso_cl}. Nevertheless, the {\small Magneticum} samples
provide cluster properties that allowed us to apply more realistic
selections.

One of the main results found by \citet{bianchi2012} is that, when
using the dispersion model, the systematic errors on the linear growth
rate are positive for DM haloes with masses lower than $\sim10^{13}
h^{-1}\,{\mbox M_\odot}$, and negative otherwise \citep[see
  also][]{okumura2011, marulli2012b}. Our findings reinforce these
conclusions. Indeed, regardless of the specific selections considered,
the overall trends of our systematic errors depend ultimately on the
mass of the DM haloes hosting the selected tracers.

Moreover, extending the \citet{bianchi2012} analysis, we investigated
how the systematic errors depend on the comoving scales
considered. Our main finding is that it is possible to substantially
reduce the systematic errors at all redshifts and selections by
restricting the analysis in a proper subregion of the
$r_\perp-r_\parallel$ plane. The latter depends on the properties of
the selected objects, specifically on the mass of the host halo, due
to the difficulties in modelling the non-Gaussian nature of the
velocity probability density function of the tracers.

%%%%%%%%%%%%%%%%%%%%%%%%%%%%%%%%%%%%%%%%%%%%%%%%%%%%%%%%%%%%%%%%%%%%%%%%%%%%%%%

\section{Conclusions}
\label{sec:conclusions}

Clustering anisotropies in redshift-space are one of the most
effective probes to test Einstein's General Relativity on large
scales. Though only galaxy samples have been considered so far for
these analyses, other astronomical tracers, such as clusters and AGN,
will be used in the next future, aimed at maximising the dynamic and
redshift ranges where to constrain the linear growth rate of cosmic
structures. In particular, clusters of galaxies can be efficiently
used to probe the low redshift Universe. Their large bias and low
velocity dispersion at small scales make them optimal tracers for
large scale structure analyses \citep{sereno2015, veropalumbo2014,
  veropalumbo2016}. On the other hand, AGN samples can be exploited to
push the GR test on larger redshifts.

In this work, we made use of both galaxy, cluster and AGN mock samples
extracted from the {\em Magneticum}, a cosmological hydrodinamic
simulation of the standard $\Lambda$CDM Universe, to investigate the
accuracy of the widely used dispersion model as a function of scale
and sample selection. Instead of analysing the clustering multipoles,
we preferred to extract $f\sigma_8$ constraints from the monopole
only, or from the full 2D anisotropic correlation \xiiz. This allowed
us to explore the $r_\perp-r_\parallel$ plane, aimed at finding the
optimal range of scales to minimise systematic uncertainties.

The main results of this analysis are the following.

\begin{itemize}

\item It is possible to get almost unbiased constraints on $f\sigma_8$
  by modelling the large-scale monopole of the two-point correlation
  function with the Kaiser limit of the dispersion model
  (Eq.~\ref{eq:xi0_2}), provided that the linear bias factor
  $b\sigma_8$ is known. The latter has to be estimated at sufficiently
  large scales, to avoid non-linearities. With real data, the linear
  bias can be estimated either with the deprojection technique, or
  from the three-point correlation function, or by combining
  clustering and lensing measurements (see the discussion in
  \S\ref{sec:methodology}).

\item When higher multipoles are considered, $f\sigma_8$ and
  $b\sigma_8$ can be jointly constrained, and the statistical error on
  $f\sigma_8$ is substantially reduced \citep[see
  e.g.][]{delatorre2013b, petracca2016}. However, systematic errors
  can arise if non-linearities are not treated accurately, such as
  with the dispersion model
  (Eqs.~\ref{eq:ximodellin}-\ref{eq:fvexp}). On the other hand, the
  analysis shown in this paper demonstrates that it is still possible
  to reduce the systematic errors on $f\sigma_8$ if the range of
  scales used to model RSD is chosen appropriately. The latter depends
  on the properties of the tracers. We do not find however any clear
  prescription to relate these scale ranges to the sample properties.

\item If the linear growth rate is estimated from the RSD of either
  galaxies or AGN, the value of $f\sigma_8$ will be underestimated at
  $z\lesssim1$, and overestimated at larger redshifts, when the fit is
  performed in the fixed range of scales $3<r_\perp,
  r_\parallel[h^{-1}\,\mbox{Mpc}]<35$. With $r_\perp^{\rm min}\sim15$
  \Mpch and $r_\parallel^{\rm max}\sim20$ \Mpch, we get almost
  unbiased constraints. However, the proper values of $r_\perp^{\rm
    min}$ and $r_\parallel^{\rm max}$ depend on sample selection, and
  should be fixed differently for each sample analysed, calibrating
  the analysis using specific mock catalogues.

\item If the RSD of galaxy groups and clusters are modelled in the
  range of scales $3<r_\perp, r_\parallel[h^{-1}\,\mbox{Mpc}]<35$, the
  value of $f\sigma_8$ results severely overestimated at all
  redshifts. The results at $z=1$ appear in overall agreement with
  what found by \citet{bianchi2012}.  The systematic error can be
  reduced if the fit is performed at larger scales, as expected due to
  the low number of cluster pairs at small separations. Specifically,
  with $r_\perp^{\rm min}\sim20$ \Mpch, we get unbiased constraints,
  but only for the densest group samples. The paucity of the most
  massive cluster samples does not allow us to obtain robust results.

\end{itemize}

To summarise, as the dispersion model cannot accurately describe the
non-linear motions at small scales, the constraints on $f\sigma_8$ can
be biased. At the present time, it is difficult to include model
uncertainties in this kind of analyses, and even more to improve the
RSD modelling at small scales \citep[see e.g.][and references
  therein]{delatorre2012, bianchi2015}. Instead of trying to improve
the model, in this analysis we investigated how the systematic errors
on the linear growth rate caused by model uncertainties depend on the
comoving scales considered.  As we have shown, in order to reduce
these systematics it is necessary to restrict the analysis in a proper
subregion of the $r_\perp-r_\parallel$ plane. More specifically, we
found that it is enough to choose a proper value for $r_\perp^{\rm
  min}$ and $r_\parallel^{\rm max}$. Indeed, as we verified, changing
also $r_\perp^{\rm max}$ and $r_\parallel^{\rm min}$ does not affect
significantly the results.

Overall, we find that, in order to reduce the systematics in the
$f\sigma_8$ measurements, it is more convenient simply to not consider
the small scales where non-linear effects distort significantly the
clustering shape, rather than to try modelling them with the empirical
description provided by the dispersion model given by
Eq.~(\ref{eq:ximodel}).  The drawback of this approach is to
inevitably increase the statistic error on $f\sigma_8$. Moreover, the
$r_\perp-r_\parallel$ region to be selected depends on the properties
of the tracers, and should be determined using suitable mock
catalogues. On the other hand, without a sufficiently accurate
modelisation of non-linear dynamics at small scales, this approach is
the only one possible to get unbiased constraints.

This can have a non-negligible impact on multiple tracer analyses
\citep[see e.g.][and references therein]{mcdonald2009, blake2013,
  marin2015} if the RSD are modelled at small non-linear scales with
the dispersion model. To avoid different systematic errors on the
linear growth rate estimated from the small-scale RSD of different
tracers, the $r_\perp-r_\parallel$ plane analysed should be different
for each sample. On the other hand, in multiple tracer techniques that
require to analyse different samples at the same scales, it is
necessary to limit the analysis at large enough scales where
systematic errors are lower than statistical errors for all the
different tracers.  And the same when joining together different
redshifts, where RSD constraints can be differently biased.

\begin{figure}
  \includegraphics[width=0.49\textwidth]{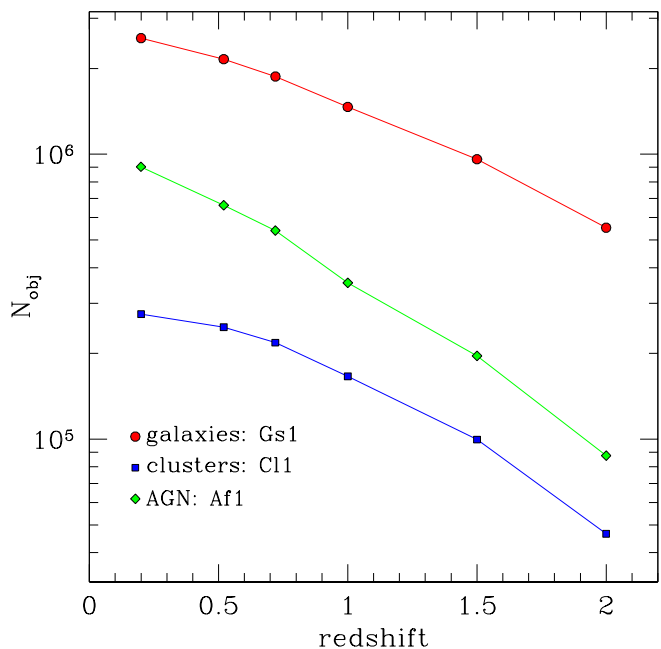}
  \caption{The number of mock galaxies ({\em red dots}), clusters
    ({\em blue squares}) and AGN ({\em green diamonds}) as a function
    of redshift, for the three selections: Gs1, Cl1, Af1 (see Tables
    \ref{tab:table1}, \ref{tab:table2} and \ref{tab:table3}).}
  \label{fig:nz}
\end{figure}

%%%%%%%%%%%%%%%%%%%%%%%%%%%%%%%%%%%%%%%%%%%%%%%%%%%%%%%%%%%%%%%%%%%%%%%%%%%%%%%

\section*{Acknowledgments}

We acknowledge the support from the grants ASI/INAF n.I/023/12/0
`Attivit\`a relative alla fase B2/C per la missione Euclid' and MIUR
PRIN 2010-2011 `The dark Universe and the cosmic evolution of baryons:
from current surveys to Euclid'. L.M. acknowledges financial
contributions from contracts and PRIN INAF 2012 `The Universe in the
box: multiscale simulations of cosmic structure'. K.D. acknowledges
the support by the DFG Cluster of Excellence `Origin and Structure of
the Universe'. We are especially grateful for the support by
M.~Petkova through the Computational Center for Particle and
Astrophysics (C$^2$PAP). Computations have been performed at the
`Leibniz-Rechenzentrum' with CPU time assigned to the Project
`pr86re', as well as at the `Rechenzentrum der Max-Planck-
Gesellschaft' at the `Max-Planck-Institut f\"ur Plasmaphysik' with CPU
time assigned to the `Max-Planck-Institut f\"ur Astrophysik'.
Information on the {\it Magneticum Pathfinder} project is available at
http://www.magneticum.org. Finally, we thank the referee for several
useful suggestions that improved the presentation of our results.

%%%%%%%%%%%%%%%%%%%%%%%%%%%%%%%%%%%%%%%%%%%%%%%%%%%%%%%%%%%%%%%%%%%%%%%%%%%%%%%

\appendix

\section{The main properties of the selected samples}
\label{app:samples}

Tables \ref{tab:table1}, \ref{tab:table2} and \ref{tab:table3} report
the main properties of the mock samples analysed in this work. We
consider only integral catalogues. The quantities {\em min} and {\em
  max} reported in the Tables are the minimum and maximum thresholds
used for the selections, while {\em median} is the median of the
sample distributions. As it can be seen, we adopt suitable sample
selections to have a minimum of $\sim50000$ objects in the smallest
galaxy samples, and $\sim10000$ objects in the smallest cluster and
AGN samples. As an illustrative example, Fig.~\ref{fig:nz} shows the
number of mock galaxies, clusters and AGN as a function of redshift,
for three different selections, as indicated by the labels.

\section{The effect of random catalogue number densities}
\label{app:random}

\begin{figure}
  \includegraphics[width=0.49\textwidth]{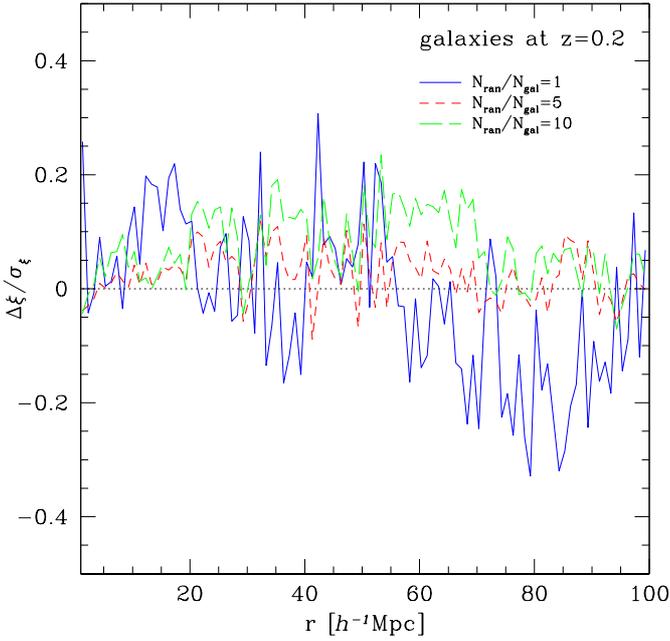}
  \caption{The ratio $\Delta\xi/\sigma_\xi$ as a function of scales
    obtained from the sample of mock galaxies at $z=0.2$, where
    $\Delta\xi=\xi_{N_R=3}-\xi_{N'_R}$, with $N_R$ being the ratio
    between the number of random objects and galaxies, and
    $\sigma_\xi$ is the estimated statistical uncertainty. Blue, red
    and green lines refer to the values of $N'_R=1, 5, 10$,
    respectively.}
  \label{fig:random}
\end{figure}

We test the impact of our assumptions in constructing the random
catalogues by repeating the full statistical analysis on a subset of
the mock samples, using random catalogues with different number
densities. Fig.~\ref{fig:random} shows an example case obtained from
the sample of mock galaxies at $z=0.2$. The lines show the ratio
$\Delta\xi/\sigma_\xi$ as a function of scales, where
$\Delta\xi=\xi_{N_R=3}-\xi_{N'_R}$, with $N_R$ being the ratio between the
number of random objects and galaxies, and $\sigma_\xi$ is the
estimated statistical uncertainty. The different lines refer to the
values $N'_R=1, 5, 10$. The effect of different random number
densities on the measured correlation functions is marginal,
considering the estimated uncertainties. As verified, the net effect
on the final RSD constraints is also negligible. 

\section{The effect of covariance matrix approximations}
\label{app:errors}

\begin{figure}
  \includegraphics[width=0.49\textwidth]{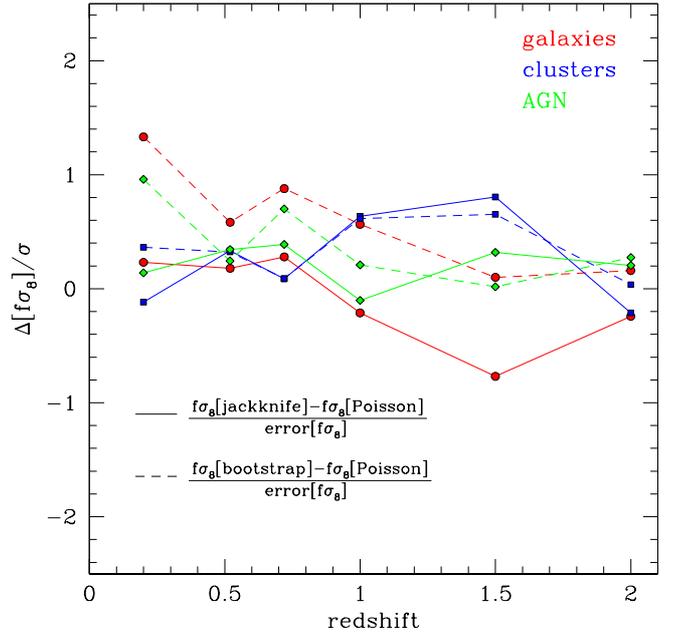}
  \caption{The difference between $f\sigma_8$ measurements obtained
    with diagonal Poisson errors and by considering the full
    covariance matrix assessed with either the jackknife (solid lines)
    or the bootstrap (dashed lines) techniques, divided by the
    $1\sigma$ uncertainty on $f\sigma_8$, as a function of
    redshift. Red, blue and green symbols refer to galaxy, cluster and
    AGN samples, respectively.}
  \label{fig:errors}
\end{figure}

All the results presented in this paper have been obtained by assuming
Poisson errors in clustering measurements and by neglecting the
off-diagonal elements of the covariance matrix. Both of these
approximations can introduce systematics in the inferred cosmological
constraints. However, as we verified, they do not impact significantly
our present conclusions, introducing just marginal differences in the
final obtained constraints, considering the estimated
uncertainties. Fig.~\ref{fig:errors} shows the difference in the
$f\sigma_8$ measurements obtained with diagonal Poisson errors and by
considering the full covariance matrix assessed with either jackknife
or bootstrap techniques, divided by the $1\sigma$ uncertainty on
$f\sigma_8$. We use $125$ jackknife/bootstrap mock realisations to
compute the covariance matrices \citep[see][for more details on the
  used methods and codes]{veropalumbo2016}. As it can be seen, the
effect of assuming negligible off-diagonal elements in the covariance
matrix is not statistically significant. Moreover, the off-diagonal
elements in the covariance matrix can introduce systematics and
spurious scatter in the inferred cosmological constraints, if not
properly estimated with a sufficiently large number of mocks.  The
above considerations motivated our choice of neglecting off-diagonal
terms in the analyses presented in this paper, in line with previous
similar works \citep[see e.g.][]{bianchi2015, petracca2016}.

\begin{table*}
  \scriptsize
  \begin{center}
    \caption{Main properties of the samples of galaxies analysed in the paper.}
    \begin{tabular}{ccrrrrrcrrrrrcrrrrrrrr}
      \hline
      \multicolumn{1}{c} {redshift} & \multicolumn{5}{c} {$\log(M_{\rm STAR} [h^{-1}\,{\mbox M_\odot}])$} & & \multicolumn{5}{c} {$M_g$} & & \multicolumn{5}{c} {$\log(SFR [{\mbox M_\odot/year}])$} \\
      \cline{2-6} \cline{8-12} \cline{14-18}
      & sample & min & max & median & objects & & sample & min & max & median & objects & & sample & min & max & median & objects \\
      \hline
      \hline
      ${\bf0.2} $ & Gm1 & $10.00$ & $13.24$ & $10.78$ & $2565412$ & & Gmg1 & $-20.00$ & $-27.39$ & $-21.90$ & $2555977$ & & Gs1 & $1.00$ & $3.40$ & $1.38$ & $1460374$ \\
      & Gm2 & $10.45$ & & $10.98$ & $1881270$ & & Gmg2 & $-20.94$ & & $-21.93$ & $2465212$ & & Gs2 & $1.24$ & & $1.47$ & $1057208$ \\
      & Gm3 & $10.90$ & & $11.13$ & $1080529$ & & Gmg3 & $-21.87$ & & $-22.31$ & $1333968$ & & Gs3 & $1.48$ & & $1.64$ & $509094$ \\
      & Gm4 & $11.35$ & & $11.55$ & $255117$ & & Gmg4 & $-22.81$ & & $-23.19$ & $267639$ & & Gs4 & $1.72$ & & $1.86$ & $169018$ \\
      & Gm5 & $11.79$ & & $11.96$ & $50000$ & & Gmg5 & $-23.74$ & & $-24.12$ & $49995$ & & Gs5 & $1.96$ & & $2.08$ & $49996$ \\
      \hline
      ${\bf0.52} $ & Gm1 & $10.00$ & $13.10$ & $10.73$ & $2167365$ & & Gmg1 & $-20.00$ & $-27.08$ & $-21.65$ & $2154603$ & & Gs1 & $1.00$ & $3.33$ & $1.45$ & $1432387$ \\
      & Gm2 & $10.42$ & & $10.93$ & $1556875$ & & Gmg2 & $-20.88$ & & $-21.70$ & $1985640$ & & Gs2 & $1.27$ & & $1.54$ & $1087917$ \\
      & Gm3 & $10.84$ & & $11.11$ & $905634$ & & Gmg3 & $-21.76$ & & $-22.21$ & $913480$ & & Gs3 & $1.54$ & & $1.71$ & $550535$ \\
      & Gm4 & $11.27$ & & $11.46$ & $245678$ & & Gmg4 & $-22.64$ & & $-23.07$ & $231625$ & & Gs4 & $1.80$ & & $1.95$ & $181393$ \\
      & Gm5 & $11.69$ & & $11.84$ & $50000$ & & Gmg5 & $-23.52$ & & $-23.88$ & $49993$ & & Gs5 & $2.07$ & & $2.20$ & $49998$ \\
      \hline
      ${\bf0.72} $ & Gm1 & $10.00$ & $13.03$ & $10.69$ & $1894509$ & & Gmg1 & $-20.00$ & $-26.78$ & $-21.53$ & $1872485$ & & Gs1 & $1.00$ & $3.35$ & $1.50$ & $1360951$ \\
      & Gm2 & $10.40$ & & $10.90$ & $1347360$ & & Gmg2 & $-20.86$ & & $-21.62$ & $1657323$ & & Gs2 & $1.28$ & & $1.58$ & $1069332$ \\
      & Gm3 & $10.81$ & & $11.10$ & $788388$ & & Gmg3 & $-21.72$ & & $-22.19$ & $708654$ & & Gs3 & $1.56$ & & $1.75$ & $569550$ \\
      & Gm4 & $11.21$ & & $11.39$ & $242502$ & & Gmg4 & $-22.58$ & & $-23.05$ & $209604$ & & Gs4 & $1.84$ & & $2.00$ & $190113$ \\
      & Gm5 & $11.61$ & & $11.77$ & $50000$ & & Gmg5 & $-23.44$ & & $-23.79$ & $49984$ & & Gs5 & $2.13$ & & $2.25$ & $49999$ \\
      \hline
      ${\bf1} $ & Gm1 & $10.00$ & $12.92$ & $10.62$ & $1507943$ & & Gmg1 & $-20.00$ & $-26.78$ & $-21.43$ & $1465209$ & & Gs1 & $1.00$ & $3.41$ & $1.58$ & $1206017$ \\
      & Gm2 & $10.37$ & & $10.84$ & $1055108$ & & Gmg2 & $-20.84$ & & $-21.55$ & $1229380$ & & Gs2 & $1.30$ & & $1.65$ & $994050$ \\
      & Gm3 & $10.75$ & & $11.07$ & $620010$ & & Gmg3 & $-21.69$ & & $-22.15$ & $490496$ & & Gs3 & $1.60$ & & $1.81$ & $572970$ \\
      & Gm4 & $11.12$ & & $11.28$ & $247210$ & & Gmg4 & $-22.53$ & & $-23.09$ & $171265$ & & Gs4 & $1.89$ & & $2.05$ & $199961$ \\
      & Gm5 & $11.49$ & & $11.65$ & $49999$ & & Gmg5 & $-23.38$ & & $-23.71$ & $49689$ & & Gs5 & $2.19$ & & $2.32$ & $50000$ \\
      \hline
      ${\bf1.5} $ & Gm1 & $10.00$ & $12.79$ & $10.55$ & $1028296$ & & Gmg1 & $-20.00$ & $-26.09$ & $-21.28$ & $959724$ & & Gs1 & $1.00$ & $3.53$ & $1.67$ & $921472$ \\
      & Gm2 & $10.33$ & & $10.76$ & $711625$ & & Gmg2 & $-20.79$ & & $-21.44$ & $760311$ & & Gs2 & $1.31$ & & $1.72$ & $811432$ \\
      & Gm3 & $10.67$ & & $10.99$ & $421555$ & & Gmg3 & $-21.58$ & & $-21.96$ & $298389$ & & Gs3 & $1.62$ & & $1.87$ & $516807$ \\
      & Gm4 & $11.00$ & & $11.19$ & $204649$ & & Gmg4 & $-22.38$ & & $-23.28$ & $86610$ & & Gs4 & $1.93$ & & $2.09$ & $206667$ \\
      & Gm5 & $11.33$ & & $11.48$ & $50000$ & & Gmg5 & $-23.17$ & & $-23.56$ & $49994$ & & Gs5 & $2.24$ & & $2.37$ & $50000$ \\
      \hline
      ${\bf2} $ & Gm1 & $10.00$ & $12.49$ & $10.48$ & $618406$ & & Gmg1 & $-20.00$ & $-25.87$ & $-21.17$ & $552844$ & & Gs1 & $1.00$ & $3.54$ & $1.74$ & $589434$ \\
      & Gm2 & $10.29$ & & $10.67$ & $425026$ & & Gmg2 & $-20.51$ & & $-21.27$ & $477671$ & & Gs2 & $1.31$ & & $1.78$ & $543876$ \\
      & Gm3 & $10.58$ & & $10.88$ & $254679$ & & Gmg3 & $-21.02$ & & $-21.48$ & $330108$ & & Gs3 & $1.61$ & & $1.90$ & $386208$ \\
      & Gm4 & $10.87$ & & $11.10$ & $130253$ & & Gmg4 & $-21.53$ & & $-21.85$ & $148563$ & & Gs4 & $1.92$ & & $2.11$ & $179827$ \\
      & Gm5 & $11.16$ & & $11.30$ & $49999$ & & Gmg5 & $-22.05$ & & $-22.49$ & $47889$ & & Gs5 & $2.23$ & & $2.36$ & $49996$ \\
      \hline
      \hline
      \label{tab:table1}
    \end{tabular}
  \end{center}
\end{table*}

\begin{table*}
  \scriptsize
  \begin{center}
    \caption{Main properties of the samples of galaxy groups and clusters analysed in the paper.}
    \begin{tabular}{ccrrrrrcrrrrrcrrrrrrrr}
      \hline
      \multicolumn{1}{c} {redshift} & \multicolumn{5}{c} {$\log(M_{500} [h^{-1}\,{\mbox M_\odot}])$} & & \multicolumn{5}{c} {$T_{500} [keV]$} & & \multicolumn{5}{c} {$\log(L_{500} [erg/s])$} \\
      \cline{2-6} \cline{8-12} \cline{14-18}
      & sample & min & max & median & objects & & sample & min & max & median & objects & & sample & min & max & median & objects \\
      \hline
      \hline
      ${\bf0.2} $ & Cm1 & $13.00$ & $14.97$ & $13.26$ & $103201$ & & Ct1 & $0.10$ & $8.27$ & $0.29$ & $274787$ & & Cl1 & $42.00$ & $46.57$ & $42.51$ & $249655$ \\
      & Cm2 & $13.18$ & & $13.41$ & $63705$ & & Ct2 & $0.19$ & & $0.35$ & $215729$ & & Cl2 & $42.41$ & & $42.75$ & $153119$ \\
      & Cm3 & $13.37$ & & $13.58$ & $36858$ & & Ct3 & $0.34$ & & $0.54$ & $110534$ & & Cl3 & $42.83$ & & $43.16$ & $65040$ \\
      & Cm4 & $13.55$ & & $13.74$ & $19963$ & & Ct4 & $0.64$ & & $0.89$ & $39922$ & & Cl4 & $43.24$ & & $43.54$ & $26922$ \\
      & Cm5 & $13.74$ & & $13.90$ & $9999$ & & Ct5 & $1.18$ & & $1.52$ & $9902$ & & Cl5 & $43.66$ & & $43.92$ & $9977$ \\
      \hline
      ${\bf0.52} $ & Cm1 & $13.00$ & $14.97$ & $13.23$ & $84523$ & & Ct1 & $0.10$ & $7.94$ & $0.30$ & $247341$ & & Cl1 & $42.00$ & $46.09$ & $42.64$ & $239836$ \\
      & Cm2 & $13.16$ & & $13.37$ & $53798$ & & Ct2 & $0.18$ & & $0.36$ & $198050$ & & Cl2 & $42.43$ & & $42.78$ & $180659$ \\
      & Cm3 & $13.31$ & & $13.50$ & $32449$ & & Ct3 & $0.33$ & & $0.52$ & $109150$ & & Cl3 & $42.87$ & & $43.19$ & $75378$ \\
      & Cm4 & $13.47$ & & $13.64$ & $18518$ & & Ct4 & $0.61$ & & $0.84$ & $40267$ & & Cl4 & $43.30$ & & $43.59$ & $29210$ \\
      & Cm5 & $13.62$ & & $13.79$ & $10000$ & & Ct5 & $1.11$ & & $1.42$ & $9959$ & & Cl5 & $43.73$ & & $43.98$ & $9991$ \\
      \hline
      ${\bf0.72} $ & Cm1 & $13.00$ & $14.69$ & $13.22$ & $69404$ & & Ct1 & $0.10$ & $6.57$ & $0.31$ & $218241$ & & Cl1 & $42.00$ & $46.24$ & $42.77$ & $215861$ \\
      & Cm2 & $13.13$ & & $13.33$ & $45804$ & & Ct2 & $0.18$ & & $0.36$ & $176278$ & & Cl2 & $42.44$ & & $42.83$ & $187288$ \\
      & Cm3 & $13.27$ & & $13.45$ & $29018$ & & Ct3 & $0.32$ & & $0.50$ & $103154$ & & Cl3 & $42.89$ & & $43.20$ & $84083$ \\
      & Cm4 & $13.40$ & & $13.57$ & $17408$ & & Ct4 & $0.58$ & & $0.80$ & $39329$ & & Cl4 & $43.33$ & & $43.61$ & $31049$ \\
      & Cm5 & $13.54$ & & $13.69$ & $9999$ & & Ct5 & $1.04$ & & $1.31$ & $9973$ & & Cl5 & $43.77$ & & $44.02$ & $9992$ \\
      \hline
      ${\bf1} $ & Cm1 & $12.50$ & $14.58$ & $12.81$ & $167773$ & & Ct1 & $0.10$ & $5.85$ & $0.31$ & $166270$ & & Cl1 & $42.00$ & $46.15$ & $42.88$ & $166445$ \\
      & Cm2 & $12.72$ & & $12.97$ & $104160$ & & Ct2 & $0.17$ & & $0.36$ & $137909$ & & Cl2 & $42.44$ & & $42.91$ & $156759$ \\
      & Cm3 & $12.95$ & & $13.15$ & $55569$ & & Ct3 & $0.30$ & & $0.47$ & $86893$ & & Cl3 & $42.88$ & & $43.20$ & $84056$ \\
      & Cm4 & $13.17$ & & $13.34$ & $25461$ & & Ct4 & $0.52$ & & $0.72$ & $36186$ & & Cl4 & $43.32$ & & $43.59$ & $32064$ \\
      & Cm5 & $13.39$ & & $13.54$ & $10000$ & & Ct5 & $0.91$ & & $1.15$ & $9991$ & & Cl5 & $43.76$ & & $43.99$ & $9992$ \\
      \hline
      ${\bf1.5} $ & Cm1 & $12.50$ & $14.43$ & $12.77$ & $100290$ & & Ct1 & $0.10$ & $4.83$ & $0.31$ & $99747$ & & Cl1 & $42.00$ & $46.20$ & $43.06$ & $100220$ \\
      & Cm2 & $12.67$ & & $12.90$ & $66549$ & & Ct2 & $0.16$ & & $0.35$ & $86504$ & & Cl2 & $42.44$ & & $43.06$ & $98730$ \\
      & Cm3 & $12.85$ & & $13.04$ & $39631$ & & Ct3 & $0.27$ & & $0.45$ & $57585$ & & Cl3 & $42.87$ & & $43.23$ & $70071$ \\
      & Cm4 & $13.02$ & & $13.18$ & $21126$ & & Ct4 & $0.44$ & & $0.61$ & $30005$ & & Cl4 & $43.31$ & & $43.60$ & $29699$ \\
      & Cm5 & $13.20$ & & $13.34$ & $10000$ & & Ct5 & $0.72$ & & $0.92$ & $9975$ & & Cl5 & $43.75$ & & $43.97$ & $9971$ \\
      \hline
      ${\bf2} $ & Cm1 & $12.50$ & $14.23$ & $12.74$ & $46857$ & & Ct1 & $0.10$ & $4.17$ & $0.28$ & $46602$ & & Cl1 & $42.00$ & $45.79$ & $43.21$ & $46857$ \\
      & Cm2 & $12.61$ & & $12.82$ & $34568$ & & Ct2 & $0.15$ & & $0.31$ & $42502$ & & Cl2 & $42.40$ & & $43.21$ & $46818$ \\
      & Cm3 & $12.73$ & & $12.91$ & $24316$ & & Ct3 & $0.22$ & & $0.39$ & $30646$ & & Cl3 & $42.80$ & & $43.23$ & $44871$ \\
      & Cm4 & $12.84$ & & $13.00$ & $15934$ & & Ct4 & $0.34$ & & $0.52$ & $18884$ & & Cl4 & $43.20$ & & $43.52$ & $23962$ \\
      & Cm5 & $12.95$ & & $13.10$ & $10000$ & & Ct5 & $0.50$ & & $0.68$ & $9934$ & & Cl5 & $43.60$ & & $43.87$ & $9953$ \\
      \hline
      \hline
      \label{tab:table2}
    \end{tabular}
  \end{center}
\end{table*}

\begin{table*}
  \scriptsize
  \begin{center}
    \caption{Main properties of the samples of AGN analysed in the paper.}
    \begin{tabular}{ccrrrrrcrrrrrcrrrrrrrr}
      \hline
      \multicolumn{1}{c} {redshift} & \multicolumn{5}{c} {$\log(M_{\rm BH} [h^{-1}\,{\mbox M_\odot}])$} & & \multicolumn{5}{c} {$\log(L_{\rm bol} [erg/s])$} & & \multicolumn{5}{c} {$\log(f_{\rm Edd})$} \\
      \cline{2-6} \cline{8-12} \cline{14-18}
      & sample & min & max & median & objects & & sample & min & max & median & objects & & sample & min & max & median & objects \\
      \hline
      \hline
      ${\bf0.2} $ & Am1 & $6.30$ & $11.05$ & $8.50$ & $946365$ & & Al1 & $40.00$ & $48.21$ & $43.26$ & $903502$ & & Af1 & $-5.00$ & $0.19$ & $-3.28$ & $900571$ \\
      & Am2 & $7.16$ & & $8.76$ & $559616$ & & Al2 & $41.49$ & & $43.35$ & $820496$ & & Af2 & $-3.78$ & & $-2.82$ & $669586$ \\
      & Am3 & $8.02$ & & $8.78$ & $530118$ & & Al3 & $42.99$ & & $43.67$ & $561712$ & & Af3 & $-2.57$ & & $-1.65$ & $282446$ \\
      & Am4 & $8.88$ & & $9.13$ & $200602$ & & Al4 & $44.48$ & & $45.26$ & $104962$ & & Af4 & $-1.35$ & & $-0.92$ & $100757$ \\
      & Am5 & $9.74$ & & $9.88$ & $9999$ & & Al5 & $45.97$ & & $46.29$ & $9999$ & & Af5 & $-0.13$ & & $0.11$ & $9999$ \\
      \hline
      ${\bf0.52} $ & Am1 & $6.30$ & $11.03$ & $8.49$ & $687068$ & & Al1 & $40.00$ & $48.27$ & $43.55$ & $662433$ & & Af1 & $-5.00$ & $0.20$ & $-2.98$ & $664774$ \\
      & Am2 & $7.15$ & & $8.84$ & $387939$ & & Al2 & $41.55$ & & $43.65$ & $602637$ & & Af2 & $-3.74$ & & $-2.75$ & $559834$ \\
      & Am3 & $8.00$ & & $8.86$ & $368515$ & & Al3 & $43.10$ & & $43.93$ & $435109$ & & Af3 & $-2.47$ & & $-1.57$ & $227878$ \\
      & Am4 & $8.85$ & & $9.12$ & $189324$ & & Al4 & $44.64$ & & $45.39$ & $86203$ & & Af4 & $-1.21$ & & $-0.69$ & $74218$ \\
      & Am5 & $9.70$ & & $9.83$ & $10000$ & & Al5 & $46.19$ & & $46.51$ & $9999$ & & Af5 & $0.05$ & & $0.18$ & $9999$ \\
      \hline
      ${\bf0.72} $ & Am1 & $6.30$ & $10.98$ & $8.18$ & $556411$ & & Al1 & $40.00$ & $48.37$ & $43.65$ & $539859$ & & Af1 & $-5.00$ & $0.20$ & $-2.80$ & $542532$ \\
      & Am2 & $7.14$ & & $8.89$ & $298180$ & & Al2 & $41.54$ & & $43.76$ & $493217$ & & Af2 & $-3.75$ & & $-2.64$ & $483058$ \\
      & Am3 & $7.99$ & & $8.92$ & $282007$ & & Al3 & $43.09$ & & $44.07$ & $351857$ & & Af3 & $-2.50$ & & $-1.63$ & $216373$ \\
      & Am4 & $8.83$ & & $9.13$ & $170321$ & & Al4 & $44.63$ & & $45.37$ & $85212$ & & Af4 & $-1.25$ & & $-0.70$ & $69112$ \\
      & Am5 & $9.67$ & & $9.80$ & $9999$ & & Al5 & $46.18$ & & $46.44$ & $9999$ & & Af5 & $0.00$ & & $0.16$ & $10000$ \\
      \hline
      ${\bf1} $ & Am1 & $6.30$ & $10.92$ & $7.99$ & $362513$ & & Al1 & $40.00$ & $48.40$ & $43.91$ & $353674$ & & Af1 & $-5.00$ & $0.19$ & $-2.63$ & $355647$ \\
      & Am2 & $7.13$ & & $9.02$ & $194441$ & & Al2 & $41.60$ & & $44.03$ & $323269$ & & Af2 & $-3.72$ & & $-2.50$ & $324605$ \\
      & Am3 & $7.97$ & & $9.05$ & $181542$ & & Al3 & $43.20$ & & $44.36$ & $235508$ & & Af3 & $-2.43$ & & $-1.46$ & $154064$ \\
      & Am4 & $8.80$ & & $9.17$ & $138296$ & & Al4 & $44.80$ & & $45.56$ & $69691$ & & Af4 & $-1.15$ & & $-0.46$ & $56167$ \\
      & Am5 & $9.63$ & & $9.75$ & $9999$ & & Al5 & $46.39$ & & $46.65$ & $10000$ & & Af5 & $0.14$ & & $0.18$ & $9959$ \\
      \hline
      ${\bf1.5} $ & Am1 & $6.30$ & $10.75$ & $7.01$ & $199546$ & & Al1 & $40.00$ & $48.52$ & $44.17$ & $196211$ & & Af1 & $-5.00$ & $0.20$ & $-2.05$ & $197164$ \\
      & Am2 & $7.12$ & & $9.14$ & $97867$ & & Al2 & $41.68$ & & $44.30$ & $181622$ & & Af2 & $-3.71$ & & $-1.90$ & $186168$ \\
      & Am3 & $7.94$ & & $9.19$ & $89539$ & & Al3 & $43.36$ & & $44.70$ & $133756$ & & Af3 & $-2.41$ & & $-1.05$ & $114389$ \\
      & Am4 & $8.75$ & & $9.27$ & $73532$ & & Al4 & $45.04$ & & $46.14$ & $49508$ & & Af4 & $-1.12$ & & $-0.29$ & $60067$ \\
      & Am5 & $9.57$ & & $9.69$ & $9999$ & & Al5 & $46.72$ & & $46.95$ & $10000$ & & Af5 & $0.17$ & & $0.18$ & $8465$ \\
      \hline
      ${\bf2} $ & Am1 & $6.30$ & $10.64$ & $6.68$ & $88522$ & & Al1 & $40.00$ & $48.63$ & $44.09$ & $87627$ & & Af1 & $-5.00$ & $0.20$ & $-1.55$ & $87913$ \\
      & Am2 & $7.08$ & & $9.13$ & $38422$ & & Al2 & $41.66$ & & $44.27$ & $82555$ & & Af2 & $-3.76$ & & $-1.50$ & $85568$ \\
      & Am3 & $7.86$ & & $9.22$ & $33150$ & & Al3 & $43.31$ & & $45.13$ & $56194$ & & Af3 & $-2.53$ & & $-1.13$ & $67129$ \\
      & Am4 & $8.64$ & & $9.32$ & $27031$ & & Al4 & $44.97$ & & $46.31$ & $30473$ & & Af4 & $-1.30$ & & $-0.40$ & $37606$ \\
      & Am5 & $9.41$ & & $9.57$ & $10001$ & & Al5 & $46.63$ & & $46.92$ & $10000$ & & Af5 & $-0.06$ & & $0.18$ & $9988$ \\
      \hline
      \hline
      \label{tab:table3}
    \end{tabular}
  \end{center}
\end{table*}

\bibliography{bib}

\end{document}